\documentclass{article}

\usepackage{arxiv}

\usepackage[utf8]{inputenc} 
\usepackage[T1]{fontenc}    
\usepackage{hyperref}       
\usepackage{url}            
\usepackage{booktabs}       
\usepackage{amsfonts}       
\usepackage{nicefrac}       
\usepackage{microtype}      
\usepackage{lipsum}		
\usepackage{graphicx}
\usepackage{natbib}
\usepackage{doi}

\usepackage{ulem}
\usepackage{cancel}
\usepackage[linesnumbered,ruled]{algorithm2e}
\usepackage{amsmath,amssymb}
\usepackage{url}
\usepackage{multirow,bigdelim}
\usepackage{booktabs}
\usepackage{array,graphicx}
\usepackage{float}
\usepackage{wrapfig}
\usepackage{lscape}
\usepackage{rotating}
\usepackage{enumitem}
\usepackage{cleveref}
\usepackage{footnote}
\usepackage{caption}
\usepackage{subcaption}
\usepackage{todonotes}
\usepackage{amsmath}
\DeclareMathOperator*{\argmax}{arg\,max}

\title{Selective Query Processing: a Risk-Sensitive Selection of System Configurations}

\author{ \href{https://orcid.org/0000-0001-9273-2193}{\includegraphics[scale=0.06]{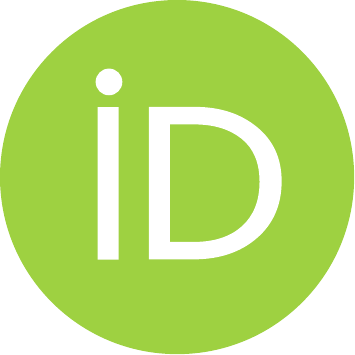}\hspace{1mm}Josiane MOTHE}\thanks{\url{http://www.irit.fr/~Josiane.Mothe)}  } \\
	IRIT, UMR5505 CNRS \\
	Universit\'{e} de Toulouse\\
	Toulouse, France \\
	\texttt{Josiane.Mothe@irit.fr} \\
	\And
	\href{https://orcid.org/0000-0002-4022-7344}{\includegraphics[scale=0.06]{orcid.pdf}\hspace{1mm} Md Zia Ullah}\thanks{now Edinburgh Napier University: Edinburgh, GB }  \\
	IRIT, UMR5505 CNRS \\
	Universit\'{e} de Toulouse\\
	Toulouse, France \\
	\texttt{zuacsea@gmail.com} \\
}

\renewcommand{\shorttitle}{Selective Query Processing: a Risk-Sensitive Approach}

\hypersetup{
pdftitle={Selective Query Processing: a Risk-Sensitive Selection of System Configurations},
pdfsubject={q-bio.NC, q-bio.QM},
pdfauthor={Josiane~Mothe, Md Zia Ullah},
pdfkeywords={Information system, Information retrieval, Adaptive information retrieval, Query driven parameterisation, Learning to rank, Search engine parameters, Risk sensitive systems},
}

\begin{document}
\maketitle

\begin{abstract}
	
In information retrieval systems, search parameters are optimized to ensure high effectiveness based on a set of past searches and these optimized parameters are then used as the system configuration for all subsequent queries. A better approach, however, would be to adapt the parameters to fit the query at hand. Selective query expansion is one such an approach, in which the system decides automatically whether or not to expand the query, resulting in two possible system configurations. This approach was extended recently to include many other parameters, leading to many possible system configurations where the system automatically selects the best configuration on a per-query basis. One problem with this approach is the system training which requires evaluation of each training query with every possible configuration. In real-world systems, so many parameters and possible values must be evaluated that this approach is impractical, especially when the system must be updated frequently, as is the case for commercial search engines. In general, the more configurations, the greater the effectiveness when  configuration selection is appropriate but also the greater the risk of decreasing effectiveness in the case of an inappropriate configuration selection. To determine the ideal configurations to use on a per-query basis in real-world systems we developed a method in which a restricted number of possible configurations is pre-selected and then used in a meta-search engine that decides the best search configuration on a per query basis. We define a risk-sensitive approach for configuration pre-selection that considers the risk-reward trade-off between the number of configurations kept, and system effectiveness. We define two alternative risk functions to apply to different goals. For final configuration selection, the decision is based on query feature similarities.  We compare two alternative risk functions on two query types: ad hoc and diversity and compare these to more sophisticated machine learning-based methods. We find that a relatively small number of configurations (20) selected by our risk-sensitive model is sufficient to obtain results close to the best achievable results for each query. Effectiveness is increased by about 15\% according to the P@10 and nDCG@10 evaluation metrics when compared to traditional grid search using a single configuration and by about 20\% when compared to learning to rank documents. Our risk-sensitive approach works for both diversity- and ad hoc-oriented searches. 
Moreover, the similarity-based selection method outperforms the more sophisticated approaches. Thus, we demonstrate the feasibility of developing per-query information retrieval systems, which will guide future research in this direction. 

\end{abstract}


\section{Introduction}
\label{Introduction}

Information retrieval (IR) systems involve several distinct component processes. These include indexing to extract the document terms that will be used for query-document matching, automatic query expansion, search weighting to decide which documents to retrieve, and ranking of retrieved documents. Each component must be set properly to optimize system effectiveness. A variety of models for each component are described in the literature. For example, the search weighting model may be BM25~\cite{robertson2009probabilistic}, language modeling~\cite{ponte1998languageLong} or another; query expansion can be implemented using Bo1, Bo2~\cite{amati2003probability}, RM3~\cite{ponte1998languageLong}, or other models. Each model has hyperparameters that influence system effectiveness and can have a variety of values. For example, $b$ and $k1$ are parameters for the BM25 search weighting model, whereas the number of terms to be added to the query is one of the parameters for query expansion models. In this paper, we will call these components and their hyperparameter values a `system configuration'. Current practice in most research search contexts, and perhaps in commercial searches also, is to select the components and hyperparameters for all the training queries or past searches only once by jointly searching and optimizing system effectiveness over the hyperparameter space~\cite{danli2018}. Hyperparameter optimization is usually done on a per-collection basis~\cite{taylor2006optimisation}. One popular method is `grid search', in which a set of possible values is defined for each parameter, then grid search determines the best value for each parameter to maximize the system effectiveness on a query set~\cite{metzler2007linear}. Alternative methods include `line search'~\cite{luenberger1984linear}, `Bayesian optimization'~\cite{li2018mergeLong}, and `transfer learning'~\cite{macdonald2015transferring}.

\begin{figure}[ht!]
\centering
\includegraphics[width=0.48\textwidth]{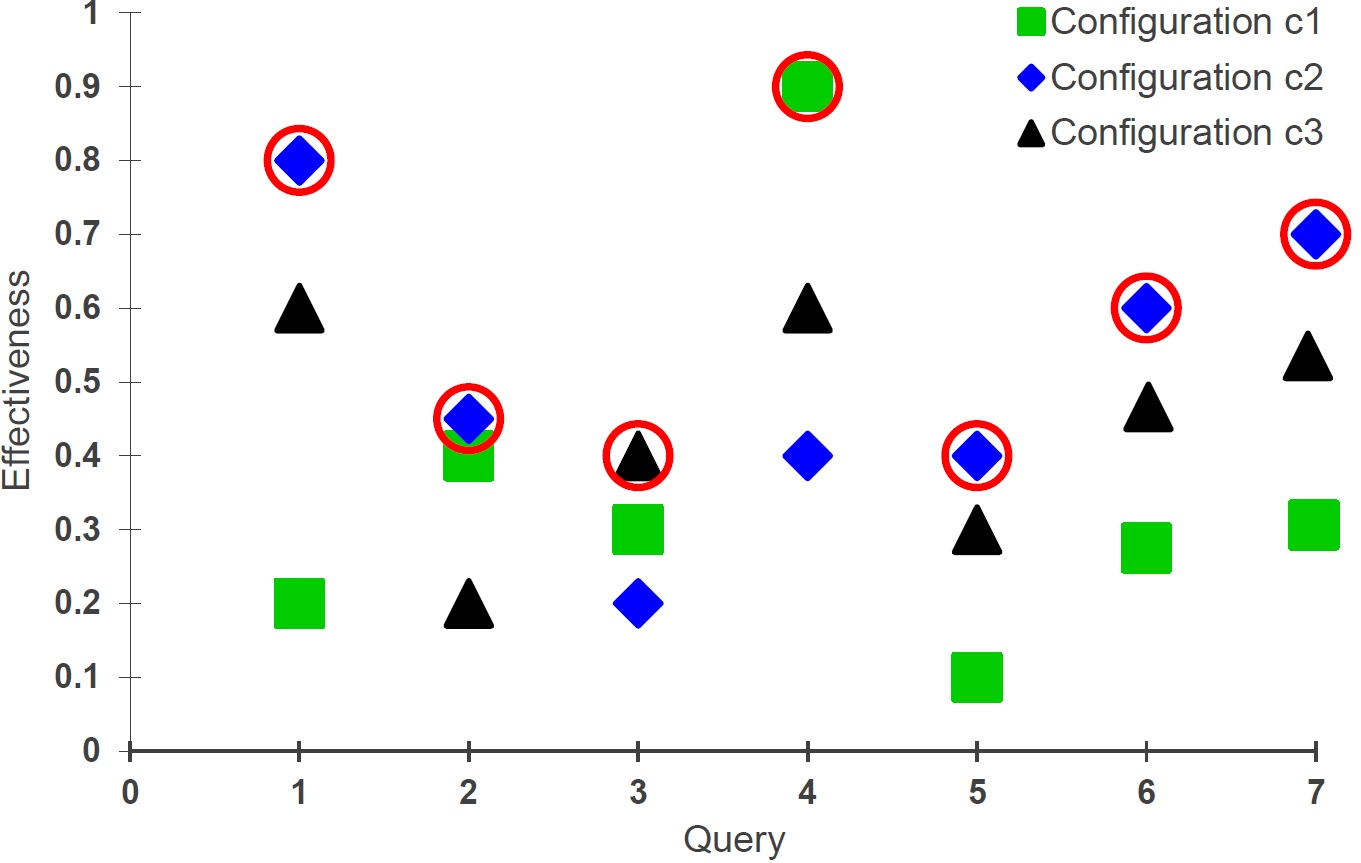}
\caption{\textbf{Various system configurations are more effective for some queries than for others}. The plot illustrates the effectiveness (y-axis) of three configurations ($c_{1}$, $c_{2}$, and $c_{3}$) for seven queries (x-axis). Whereas $c_{2}$ is best for most queries, $c_{1}$ is best for query 4, and $c_{3}$ is best for query 3. If a single configuration was to be used for all queries, then $c_{2}$ should be selected. The best possible effectiveness, however, would be to select the best configuration for each query (red circles).}
\label{fig:Illustration}
\end{figure}

The hyperparameter optimization approaches described above assume that it is possible to select a single system configuration that will fit any query. In reality, this is false: the optimal configurations for various queries differ. This has been demonstrated by the TREC\footnote{TREC: Text REtrieval Conference \url{trec.nist.gov}}, where, for example, system A performs better than system B on certain queries whereas system B performs better than system A for other queries~\cite{harman2009overview}. One query may need extensive query expansion, for example, whereas another does not. Ideally, query expansion parameters should be set differently for different queries. The same is true for other system components and hyperparameters. Even if a single system configuration could optimize effectiveness on average over a set of queries, other configurations may be more effective for each individual query. It would be better, therefore, to adapt the hyperparameter values to fit each given query (Figure~\ref{fig:Illustration}). 

Data fusion may be an alternative to selecting the best configuration for each query. It uses different configurations and then merges the results based on document ranks or scores~\cite{fox1993combining}. Potentially, data fusion might perform better than the single best configuration selected by an Oracle, however, it would require more time to compute the ranking across configurations for each query and more complex calculations.

In this paper, we will consider how to determine a hopefully good configuration for each query without taking into account the complexity and time required for computing. Previous work made some attempts in this direction. Selective query expansion (SQE), for example, tries to determine whether or not a query needs to be expanded; it is used only for those queries that need to be expanded~\cite{cronen2004framework,Cao2008prf,Xu2016}. The use of only two search strategies in SQE does not improve search engine performance very much, however, thus it is no longer used.

Recently, the SQE idea has been expanded to include many other system parameters, including indexing schema, retrieval models (BM25, language model, etc.) and the feedback methods~\cite{BigotDM15,mothe2017predicting,deveaud2018learningLong,arslan2019selective,MotheCIKM2021}. When a large number of possible parameters are involved they may be mutually dependent. For example, the number of terms selected for query expansion may depend on the number of feedback documents from which the terms are selected. It is difficult to optimize the selection of each parameter value separately~\cite{mothe2017predicting}. Also, following on Ferro and Silvello' deep analysis of the influence of system parameters on results~\cite{ferro2016}, we previously defined a large set of possible system configurations that use different parameter values~\cite{deveaud2018learningLong}. This creates many possible system configurations that the engine must choose between to best fit a given query. A model is then trained to select the best one among these configurations depending on the query. In this way, the parameters can influence each other within the same configuration and the problem of dependencies among parameters can be tackled implicitly. The model is trained using learning to rank: query features are extracted from a given query and the model learns to rank the configurations in order to select the best one~\cite{deveaud2018learningLong,MotheCIKM2021}. The assumption in this case is that a specific configuration can best cope with the specific needs of a given query.

To implement this method, each training query must be evaluated with every configuration to obtain a measure of the effectiveness of that configuration, which serves as the ground truth. Even though the learning to rank model trained in this way can outperform previous state-of-the-art methods, the cost of preparing the training data is huge: about $20,000$ configurations were used by Deveaud \textit{et al.}~\cite{deveaud2018learningLong}, but, in practice, many more may be needed and some of the parameters have continuous values. This approach is thus difficult to use in practice, especially when the number of training queries is large. If we limit the number of training queries to make the approach more feasible, however, we face the problem of model overfitting: the selected configurations may tend to fit a few training queries, but are hard to generalize to other queries.

In our recent work~\cite{MotheCIKM2021}, we hypothesized that it is not necessary to consider all the possible configurations all of the time. Some may be very similar to others and will produce similar results, so they might be considered redundant. Specific setting of some parameters may also not be important in some contexts given the setting of other parameters. Indeed, Compaor\'{e} \textit{et al.}~\cite{compaore2011mining} and Ferro and Silvello~\cite{ferro2016} showed that some parameters have a much higher impact than others. If it were feasible to reduce the large number of possible configurations to a much smaller set of representative configurations that work well for different queries in different situations, the configuration selection approach would then be usable in practice, especially for Web search engines, which need to be updated regularly. This paper extends our previous work~\cite{MotheCIKM2021} by proposing a formal method to select a small set of system configurations from the many configurations such that for each query one configuration can be found that produces the best or near-best effectiveness.

The problem is not trivial. Consider again the case in Figure~\ref{fig:Illustration}. Suppose that we want to limit the number of candidate configurations to a subset of, say, two configurations. Since $c_2$ is best for many queries, it should be kept as a first choice. For second configuration, either $c_1$ or $c_3$ would improve effectiveness by one query, but the gain from choosing $c_1$ for query $q_4$ is much higher than that from choosing $c_3$ for query $q_3$. Therefore, $c_1$ is a better choice, even though the average effectiveness of $c_3$ on all the queries is higher than that of $c_1$. For the subsequent choices, it is important to consider whether a configuration is complementary to those already selected. This example shows that we need an optimizer to decide which configurations to keep.

In deciding which configurations to keep, one consideration is that the more configurations we keep (up to a certain point), the greater the effectiveness. Another is that the more diverse the configurations in terms of their optimal treatment of some types, the better. The third consideration is that at a certain point adding configurations leads to overfitting, specifically if there are few training queries. We propose a `greedy' approach in which we add one representative configuration at a time according to the above principle. The added configuration must contribute to improving global performance, i.e., the number of queries that improved effectiveness or the overall improvement in effectiveness.

Returning to our example in Figure~\ref{fig:Illustration}, if we use a single configuration for all the queries, using $c_{1}$ rather than $c_{2}$ for query $q_{1}$ is risky because it will lead to lower effectiveness. The overall risk of using $c_{1}$ rather than $c_{2}$ can be evaluated by considering all the queries and calculating the accumulated risk for all the individual queries. If, instead, we use $c_{3}$, the risk will be different. Risk can be estimated in several ways but, for an individual query, it is the amount we can expect to decrease retrieval quality by using one configuration over a reference configuration~\cite{wang2012robust,dinccer2016risk,benham2017risk}.

The configuration selection method we propose here is based on sensitivity to risk. Inspired by the risk-sensitive learning to rank approach, which aims to reduce the risk of poor effectiveness for some queries~\cite{wang2012robust,collins2014trec,dinccer2016risk,sousa2016incorporating}, we suggest a risk-based criterion to decide which configurations should be included as candidate in the system. The risk is defined as the loss of performance when choosing a particular system compared to the baseline. Our model thus combines risk and reward functions to select a small set of candidate configurations. The resulting reduced set of configurations can then be used by a meta-search engine to select the best configuration for a new query. 

In our previous work~\cite{MotheCIKM2021}, we considered a risk-reward function that accumulates the risk (or reward) relative to queries in terms of effectiveness when applied to the training query set, i.e., the size of the decrease (or increase) in system effectiveness due to adding a given configuration. Here, we consider an alternative risk-reward function that focuses instead on the number of queries that may be affected by a decrease (or increase) in system effectiveness. This new function is more user-oriented than the previous one, which was oriented towards system evaluation and, therefore, likely to work well in evaluation campaigns like TREC. Another novelty compared to earlier related work is how we select the best configuration for a given query. While related work, including ours, has considered complex machine learning models, here we use a simpler method: given a new query and its features, we find the configuration associated with the best-matched training query and use it for the underlying query. Thus, the similarity based on the query features only is used. 

To evaluate our method, we consider six reference test collections from TREC. We extend our previous analysis to the diversity task and the ad hoc task and include the MS MARCO data collection. We compare our method with state-of-the-art methods, including grid search to select the best single configuration, learning to rank methods for document ranking, selective query expansion, data fusion (CombSUM), and Oracle. Our analyses show that the proposed method can significantly reduce the number of candidate configurations while retaining a similar level of overall effectiveness as when using many more configurations, making it usable in practice. We show empirically that our method is superior to state-of-the-art approaches on both ad hoc retrieval and diversity test collections, considering many collections, baselines and related work.

The rest of the paper is organized as follows: Section~\ref{sec:related} presents the related work. In Section~\ref{sec:risk}, we introduce the risk-based criteria to select the candidate configurations as a preliminary step of any per-query configuration selection model. Section~\ref{sec:eval} describes the evaluation framework, and Section~\ref{sec:res} reports the results and discusses them. Section~\ref{sec:conclusion} concludes the paper and provides some perspective.

\section{Related work}
\label{sec:related}

The related work falls into two categories: methods that aims at using selective query processing strategy on a per-query basis and risk-sensitive criteria in learning to rank.

\subsection{Selective query processing}
In information retrieval, selective strategies have been introduced either to increase efficiency or effectiveness. Selective search mainly refers to techniques to improve the response time by limiting the part of the data the system is searching from, like in shard selection, for example~\cite{baeza2009efficiency,aly2013taily}. On the other hand, other selective approaches have been introduced that focus on different aspects of the query processing. For example, selective query expansion was introduced to name a system that decides whether a query should be automatically expanded or not before the system retrieves the documents. Our paper focuses on this second type of selective approaches, so do this related work section.

A few studies explored the idea of applying different processes according to queries. 
A selective query processing aims to select the most appropriate search or system configuration among many to apply to each individual query~\citep{he2004query}; it thus differs from \textit{data fusion}, which fuses the retrieved document lists obtained by several system configurations~\cite{fox1994combination}, but still uses a single strategy to process all the queries~\cite{fox1994combination,nuray2006automatic,hsu2005comparing,kurland2018fusion}. It also differs from \textit{double fusion}, which fuses the retrieved document lists over query variations for a single system (query fusion) and over multiple systems for a single query (system fusion)~\cite{mccabe2001system,Bailey,benham2017risk}. Finally, it differs from \textit{distributed selective search}, which aims at avoiding to search the entire corpus for each query~\cite{mohammad2018dynamic}. 

\textit{Selective query expansion} (SQE) is probably the first approach toward selecting the process to apply for each individual query~\cite{amati2004query,cronen2004framework,he2007combining,zhao2012automatic}. For each query, the SQE approach selects one of two query processing strategies or system configurations: either the original, unexpanded query is used or the query is expanded first. Most studies on SQE focused on how this decision is made. Cronen-Townsend \textit{et al.}~\cite{cronen2004framework} and Amati \textit{et al.}~\cite{amati2004query} estimated how much the results of the expanded query strayed away from the sense of the original query to decide whether it should be expanded. 
Amati \textit{et al.}~\cite{amati2004query} reports that an Oracle that could decide which queries would need expansion and which would not make an increase of about 10\% to 13\% on MAP and P@10 compared to a single configuration without expansion (on early TREC collections). About the same percentage was reported in Cronen-Townsend \textit{et al.}~\cite{cronen2004framework} (+7\% for TREC7 and +11\% on TREC8). This considers an Oracle could decide, and this Oracle could not be achieved automatically. 

The two configurations used in SQE, however, had only limited effectiveness. The following attempts focused on query expansion since query expansion has shown to be effective on average, specifically for difficult queries~\citep{compaore2011mining}. 
Xu \textit{et al.}~\cite{xu2009query} improved SQE such that the system chooses between more options and consider different query expansion strategies according to the type of query. They defined three types of query: ambiguous queries, queries about a specific entity and broader queries, and proposed different methods to treat each type of queries. Kevyn Collins-Thompson~\cite{collins2009reducing} introduced a constraint optimization framework for SQE by reducing the risk of query expansion without hurting the average gain.

Other attempts to use a selective query processing strategy have considered more than two configurations. Arslan and Din\c{c}er~\cite{arslan2019selective} used eight configurations consisting of eight term-weighting models in which the hyperparameters were first optimized by using grid search~\cite{taylor2006optimisation}. In their approach, the selection of a configuration to process a new query is based on the  distributions of the frequency of query terms on the document collection. By considering more configurations in this selective  strategy, the system performance improved when compared to SQE~\cite{xu2009query,arslan2019selective}. Din\c{c}cer \textit{et al.}~\cite{arslan2019selective} applied the selective approach to 11 term weighting approaches. They reported a potential increase of 34\% (oracle) when compare to the best system of the pool but achieved +3\% only with their model on Clueweb09A. 

The most recent selective query processing strategies make the configuration selection decision based on training by machine learning from sample queries~\cite{he2004query,Xu2016,mothe2017predicting,deveaud2018learningLong} rather than making the decision based on a query score or feature as was done previously~\cite{cronen2004framework,xu2009query,arslan2019selective}. He and Ounis~\cite{he2004query} introduced a model selection approach in which queries were clustered considering three pre-retrieval features and the best-performing retrieval model was attributed to each cluster. For a given query, the retrieval model attributed to the closest cluster was then selected to treat the query. The same group of authors reported an increase of about 3\% on WT10G compared to a unique optimized configuration but did not report the Oracle in their selective query expansion model based on document fields~\cite{he2007combining}.

The per-parameter learning (PPL) method by Mothe and Washha~\cite{mothe2017predicting} predicts the best configuration to use for a query by considering independently each of the query processing component and hyperparameters. PPL trains a multi-class classifier for each component process and hyperparameter.  
Seven such classifiers were trained corresponding to more than $80,000$ configurations in total. These classifiers were then used to predict the best component and hyperparameter values for an unseen query, considering the query features. An independently trained classifier for each parameter may not effectively model system performance since the parameters might be mutually dependent~\cite{ferro2016}. In this paper, compared to the PPL, we combine all the components and hyperparameters in a single configuration and generate different configurations by varying the components and their hyperparameters. In this way, interaction among components and hyperparameters is allowed within the configurations.

Xu \textit{et al.}~\cite{Xu2016} focused on the automatic query expansion component of selective query processing strategy. They adapt the standard \textit{learning to rank} documents model whose purpose is to rank a sample of documents retrieved by a search system according to their supposed relevance by learning from examples of query-document preferences~\cite{trotman2005learning,li2011short}. The adaption was that the model does not learn document ranking but, rather, it learns to rank the candidate expansion terms based on query-term preferences. In Deveaud \textit{et al.}~\cite{deveaud2016learning,deveaud2018learningLong}, the configuration selection for new queries was also cast as a problem of learning to rank. Their model learns to rank a set of configurations according to their potential ability to retrieve relevant documents for a given query based on examples of query-configuration preferences. They considered various components and hyperparameters and built more than $20,000$ configurations. Although this approach has been shown to be effective, it is not applicable in practice because the huge number of configurations it considers is too costly in terms of computing time and maintenance. None of the related work considers the problem of the choice of the candidate configurations in selective query processing strategy. In many of the previous studies just a few configurations were considered but their choice was not argued. When there were too many to be handled, they were selected randomly.

In this paper, we solve this problem with a risk-reward-based method that pre-selects a manageable number of the candidate configurations. A commercial search engine cannot maintain thousands of search configurations or thousands of query reformulations and search components, specifically because they also need to be regularly tuned according to the evolution of queries. Our method requires that only a small, manageable number of configurations be maintained. Once the set of candidate configurations is obtained, then any selective query processing strategy can be applied.

\subsection{Risk-sensitive criteria in L2R}

In information retrieval, risk has been defined as the reduction in effectiveness of a system when compared to a given baseline system ~\cite{wang2012robust,Dincer:2014,dinccer2016risk,benham2017risk,benham2019pluses}. Risk-sensitive functions have been studied in the context of {learning to rank} documents where a document ranking function is learned. De Sousa \textit{et al.}~\cite{sousa2019risk} applied it to feature selection and show their models got a good trade-off between effectiveness and robustness. 
In this model, however,  both this learned function and the search strategy are the same for any query.

Inspired by the modern portfolio theory, Wang and Zhu~\cite{wang2009portfolio} presented a risk-averse ranking algorithm by considering the mean-variance analysis of a ranked list. Wang \textit{et al.}~\cite{wang2012robust} defined $F_{risk}$, which estimates the average reduction in effectiveness by using a given document ranking model rather than a reference model. Also, they proposed a risk-reward trade-off function, $U_{risk}$, to directly optimize a \textit{risk-sensitive learning to rank model} that lowers the risk and enhances the reward of the document ranking model performance. The authors use this principle to select the best unique ranking model to be applied to any query. They show that risk-sensitive optimization for ranking models leads toward both robust and effective models.

The problem of $U_{risk}$ is that it is unclear whether the estimated loss of a system over a baseline is statistically significant~\cite{Dincer:2014}. Other variants of risk-reward trade-off functions have been proposed in the literature based on $F_{risk}$~\cite{Dincer:2014,dinccer2016risk,benham2019pluses}  to overcome the problem of $U_{risk}$. Din\c{c}er \textit{et al.}~\cite{Dincer:2014} introduced $T_{risk}$, an inferential version of $U_{risk}$ that follows a Student's t-distribution. The estimated risk-reward trade-off of a system over a baseline {is}  tested in terms of its statistical significance, based on $T_{risk}$. Later, Din\c{c}er \textit{et al.}~\cite{dinccer2016risk} also proposed $Z_{risk}$ and $G_{risk}$ to compare the risk-reward trade-off of a system against multiple baselines.

As an interesting application of risk functions, De Sousa \textit{et al.}~\cite{sousa2016incorporating} incorporated risk functions for feature selection in \textit{learning to rank} documents. They compared the models obtained by considering different subsets of features in order to select the most important ones. Benham \textit{et al.}~\cite{benham2017risk} introduced the risk function to estimate the risk-reward trade-off in rank fusion. Later, Benham \textit{et al.} tried an S-shaped weighting function instead of the linear weighting function in risk measures; they found no conclusive differences in risk sensitivity~\cite{benham2019pluses}. They also suggested a naming convention and a reversed signed version of the existing risk measures so that higher values correspond to higher risk.  Benham \textit{et al.}~\cite{benham2019taking} studied the inferential behavior of risk measures and the stability of its confidence intervals. They found that the distribution of risk-adjusted scores is asymmetrical, undermining the normality assumption of the t-test statistic used in $T_{risk}$. Therefore, the inferential statistic on $T_{risk}$ is not stable.

In the approach described below, we develop  {risk-sensitive criteria that aim to determine which configurations should be included in a pool for a selective query processing strategy, the candidate configurations.} 
We develop two variants of  penalty and reward functions; one is based on the overall system effectiveness whereas the other considers the number of queries that are improved/degraded. The former considers cumulative effectiveness {and thus is likely to be more appropriate to perform well in shared tasks};  the latter considers cumulative number of queries {and thus is more likely to satisfy a larger number of users}.

\section{Risk-based criteria to select candidate configurations}
\label{sec:risk}
\begin{figure}[ht!]
\centering
\includegraphics[width=0.99\textwidth]{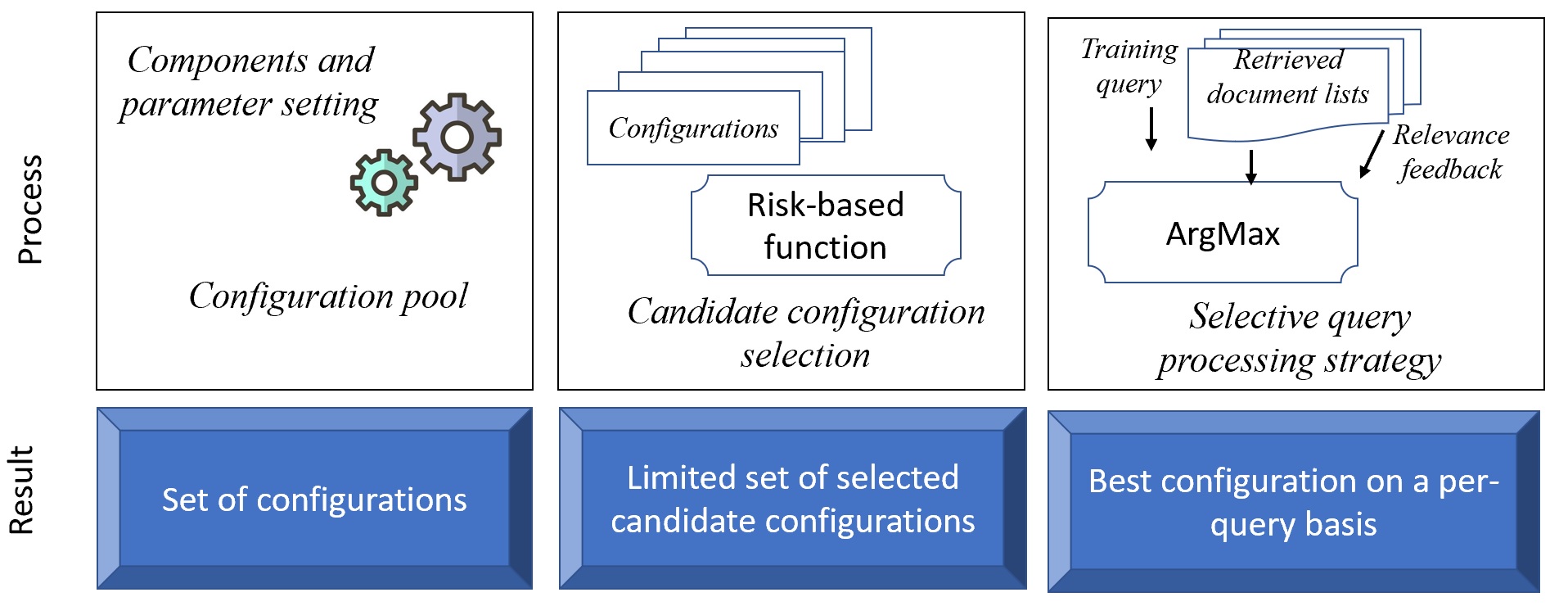}
\caption{{The training phase of model is composed of three phases: (1)  configurations are generated using different values of the query processing parameters, (2) the set of configurations is restricted using a risk-reward criterion considering the training queries to obtain the candidate configurations for the next step, and (3) the best configuration is selected on a per-query basis for each training query}.}

\label{fig:Overview}
\end{figure}

The overview of the model training for our risk-sensitive selective approach is presented in Figure~\ref{fig:Overview}. The model is trained to automatically decide which system configuration is the most appropriate for a new query.

This {model training}  consists into three main phases as follows:
\begin{enumerate}
    \item Configuration  generation: on the server point of view, a search relies on various components or modules (indexing, searching, document scoring and ranking), each implying many hyperparameters. The configuration generator creates system configurations; each configuration is able to run any query and can provide a ranked list of documents it considers to be appropriate to deliver. Depending on the query, some configurations will be more successful than others. The number of configurations can be huge considering the variety of the different components and their hyperparameters that have been suggested in the literature;
    \item Candidate configuration selection: this second phase consists in reducing the above mentioned set of  configurations. Indeed, some configurations may be poor in terms of effectiveness whatever the query is; some configurations may be very close one to the other. The risk-reward-based function for candidate configuration selection has been defined for this purpose. This phase is crucial to  make the model applicable in real-world environments but also to avoid overfitting when the number of example queries is limited.
    \item Query processing selection for training queries: the unique configuration from the candidate set that should treat a query is decided. It is based on the configuration effectiveness on that query.

\end{enumerate}

{After training, the model is able to decide which configuration it should choose to process a new query. The most similar training query based on query features is estimated; then the model   selects the configuration that was associated with it during the training.} 
In the rest of the paper, we use \textit{meta-system} or \textit{meta-search engine} when referring to the retrieval that selects a configuration on a per-query basis. \textit{Configuration} thus refers to a specific setting of the combination of modules along with the value of their hyperparameters (e.g., BM25 using the default values of $k_{1}$ and $b$, with no query expansion); A configuration is a query processing strategy.   And \textit{candidate configurations} are the configurations the meta-system  can choose among for processing a given query.  

The rest of this section is devoted to the detailed description of this model.

\subsection{{Configuration pool}}
\label{subsec:config}

In IR evaluation, it is usual to tune the system components and their hyperparameters in order to optimize the system effectiveness; once done the obtained configuration is used for all the queries. System components are the retrieval model used in the search process (e.g., LM, BM25, etc.), the query expansion model (e.g., Bo2), the re-ranking algorithm used, etc. Their influence on the system effectiveness has been studied in various work~\cite{compaore2011mining,BigotDM15,ferro2016}. The hyperparameters of those components are the $k_{1}$ and $b$ parameters for BM25 model, the $\mu$ for the LM model, etc. Lv and Zhai~\cite{lv2011documents} and Trotman \textit{et al.}~\cite{Trotmanbm25lm} for example studied the case of BM25 inner parameters.

The various configurations are built by varying them. 
The framework we propose later in this paper can be applied at the component level, hyperparameter level, or both component and hyperparameter levels. In the evaluation part of the framework, we will detail the components and hyperparameters we considered in the more than $20,000$ configurations we used.

\subsection{Risk functions to choose the candidate configurations}
\label{subsec:risk}
A crucial issue to make query-based configuration selection applicable in real systems is the number of candidate configurations to be embedded in the meta-system. 

Obviously, the more configurations, the more costly the meta-system. Moreover, we hypothesize that the more candidate configurations the meta-system can choose among (i.e., the larger the space of possible configurations), the greater the overall model effectiveness (in case of appropriate selection) but also the greater the risk (in case of bad selection of). To solve the problem of effectiveness/cost trade-off, we need an optimized number of complementary configurations in the meta-system.

Configurations used by the meta-system can be selected randomly from a set of candidate configurations using random search~\cite{bergstra2012random} as it is done in our previous work~\cite{deveaud2018learningLong}, but we believe that a risk-based model can be better. 

In this paper, we developed a theoretical framework leading to a principled process, starting from one effective configuration and adding complementary effective configurations. Intuitively, a straightforward choice for the first configuration to select could be the configuration that maximizes the average effectiveness over the training queries ($c_2$ in our toy example). Now, if we would like to select a second configuration, then it should be chosen as complementary to the first one: it should be more effective than the first configuration at least on a few queries (it can get a reward for that) and at the same time, it should not hurt many queries (bring risk into the system) in case the selection model chooses it by mistake for a given query. The process of selecting the candidate configurations can be modeled as a trade-off between minimizing the risk if the meta-system does not pick-up the best candidate configuration and maximizing the reward if the meta-system selects the best candidate configuration for a given query.

Our starting point is the risk-based approach $F_{Risk}$ suggested by Wang \textit{et al.}~\cite{wang2012robust} for the case of L2R documents and which is defined as follows:
\begin{equation}\label{eq:F}
F_{Risk}(\mathcal{Q}, M)=\frac{1}{|\mathcal{Q}|} \sum_{q_{i} \in \mathcal{Q}} \max (0, B(q_{i})- M(q_{i}))
\end{equation}

where $\mathcal{Q}$ is the training query set, $B(q_i)$ is the baseline effectiveness for query $q_{i} \in \mathcal{Q}$, and $M(q_{i})$ is the effectiveness of the  document ranking model for which risk is estimated. 

The $F_{Risk}$ thus calculates the average decrease over the training queries in terms of effectiveness when comparing two document ranking models or a baseline and a ranking model. Inspired by this $F_{Risk}$ function, we define two variants that measure the risk associated with selecting, on a per-query basis, the configuration $c_{j}$ rather than the reference configuration $c_{r}$. The risk is here to have $c_{r}$ better than $c_{j}$.

We define two risk-sensitive functions that result in two gain functions, both consisting in a risk and a reward parts. 
The first risk-sensitive function, $E_{Gain}$ aims at optimizing the overall effectiveness on the training query set; it measures risk and reward in terms of cumulative effectiveness over queries. The second risk-sensitive function, $N_{Gain}$ aims at optimizing the number of queries for which the effectiveness is improved. 

\vspace{6pt}
\noindent\textbf{Effectiveness-based function.}
For effectiveness, Eq. \ref{eq:F} can be transposed  into :
\begin{equation}\label{eq:riskEff}
 \frac{1}{|\mathcal{Q}|} \sum_{q_{i} \in \mathcal{Q}} \max(0, p(c_{r}, q_{i}) - p(c_{j}, q_{i}))
\end{equation}
where $\mathcal{Q}$ is the training query set, $p(c_{r}, q)$ is the performance (effectiveness) of the reference configuration $c_{r}$ for the query $q$, and $c_{j}$ is the candidate configuration.  

It accumulate the risk, that is to say the decrease in effectiveness, due to $c_j$ being a lower performing configuration than the reference configuration $c_j$ over the queries.

The corresponding reward function is based on the potential increase of the overall effectiveness on the training query set and is defined in Eq.~\ref{eq:rewardEff} as:
 \begin{equation}\label{eq:rewardEff}
 \frac{1}{|\mathcal{Q}|} \sum_{q_{i}\in \mathcal{Q}} \max(0, p(c_{j}, q_{i}) - p(c_{r}, q_{i}))
\end{equation}
It aggregates the effectiveness improvement that would result if the system considers $c_{j}$ for the queries where $c_{j}$ performs better than the reference configuration $c_{r}$.

Once risk and reward are defined, we have to adapt the above formulas to fit the problem of selecting a set of candidate configurations to be used in the per-query configuration selection meta-system.

Let $\mathcal{R}$ be the set of all candidate configurations, $S_{k-1}$ be the set of configurations that have already been selected at the step $k$ with an initial $S_{0} = \{ \}$.
$\mathcal{Q}$ is the set of training queries, $p\;(c_{k}, q_{i})$ denotes the retrieval effectiveness (e.g. ERR-IA@20) for the query $q_{i} \in \mathcal{Q}$ processed by the configuration $c_{k}$.

Given $\mathcal{R}$ and $S_{k-1}$, we define the risk for selecting the new configuration $c_{k}$ to be added in $S_{k-1}$ at step $k$ using Eq.~\ref{eq:riskEff} in terms of effectiveness as follows: 
\begin{equation}\label{eq:RISK_eff}
E_{Risk} (c_{k}, S_{k-1}) = \frac{1}{|\mathcal{Q}|} \sum_{q_{i}\in \mathcal{Q}} \max(0, \max_{c_{j}\in S_{k-1}} (p(c_{j}, q_{i})) - p(c_{k}, q_{i}))
\end{equation}
where $\max_{c_{j}\in S_{k-1}} p(c_{j}, q_{i})$ is the maximum effectiveness for the query $q_{i}$ over the set of configurations that have already been selected in $S_{k-1}$. In Eq.~\ref{eq:RISK_eff}, the risk for adding the configuration $c_{k}$ is measured as the cumulative decrease in effectiveness the meta-system can get if it chooses $c_{k}$ rather than the best configuration in $S_{k-1}$ for each of the training queries; it thus adapts Eq.~\ref{eq:riskEff}. The value is in between 0 and 1 when the effectiveness measure is in between 0 and 1.

Likewise, we define the  reward function using Eq.~\ref{eq:rewardEff} regarding the effectiveness as follows:
\begin{equation}\label{eq:REWARD_eff}
E_{Reward} (c_{k}, S_{k-1})= \frac{1}{|\mathcal{Q}|} \sum_{q_{i}\in \mathcal{Q}}\max(0,  p(c_{k},q_{i}) - \max_{c_{j}\in S_{k-1}} p(c_{j},q_{i}))
\end{equation}

The overall gain for the configuration $c_{k}$ over the set of training queries and already selected configurations $S_{k-1}$ is defined as:
\begin{equation}\label{eq:gain_eff}
E_{Gain}(c_{k}, S_{k-1}) =  E_{Reward}(c_{k}, S_{k-1}) - (1+\beta) E_{Risk}(c_{k}, S_{k-1})
\end{equation}
$\beta $ is a risk-sensitive parameter that controls the trade-off between risk and reward. Here, we set $\beta =0$  to weight risk and reward equally. We keep the analysis of this risk-sensitive parameter for future work. {This aspect has been considered in other contexts}~\cite{dinccer2016risk,benham2019taking}.

\vspace{6pt}
\noindent\textbf{Number of impacted queries based function.}
For $N_{Gain}$, the risk is defined in terms of the number of queries  for which the query effectiveness would decrease if the candidate configuration $c_{j}$ was selected in place of the reference configuration $c_{r}$. In that case, Eq. ~\ref{eq:F} can be transposed into :
\begin{equation}\label{eq:riskNbQ}
 \frac{1}{|\mathcal{Q}|} \sum_{q_{i}\in \mathcal{Q}} | p(c_{r}, q_{i}) - p(c_{j}, q_{i}) > 0 |
\end{equation}
where $|a > 0| = 1$ when the condition is true and using the same notation as previously. This is the maximum possible risk. It is a proportion of queries.

The corresponding reward function  corresponds to the potential number of queries that can be improved using configuration $c_{j}$ (Eq.~\ref{eq:rewardNbQ}). It  is defined as:
\begin{equation}\label{eq:rewardNbQ}
\frac{1}{|\mathcal{Q}|} \sum_{q_{i}\in \mathcal{Q}} |  p(c_{j}, q_{i}) - p(c_{r}, q_{i})  > 0 |
\end{equation}
where we count the number of queries when configuration $c_{j}$ performs better than the reference configuration $c_{r}$.

We  define the  risk function which adapts Eq.~\ref{eq:riskNbQ} to the selective meta-system by focusing on the number of queries which degrade the effectiveness as follows:
\begin{equation}\label{eq:RISK_nbq}
N_{Risk} (c_{k}, S_{k-1}) = \frac{1}{|\mathcal{Q}|} \sum_{q_{i}\in \mathcal{Q}} |(\max_{c_{j}\in S_{k-1}} p(c_{j}, q_{i}) - p(c_{k}, q_{i}) )>0|
\end{equation}
In Eq.~\ref{eq:RISK_nbq}, the risk for adding the configuration $c_{k}$ is measured as the number of queries in the training query set for which the effectiveness is lower for $c_{k}$ than for the best configuration in $S_{k-1}$.

We also define the alternative reward function using Eq.~\ref{eq:rewardNbQ} in terms of the number of queries for which the configuration gets the best effectiveness as follows:
\begin{equation}\label{eq:REWARD_nbq}
N_{Reward} (c_{k}, S_{k-1}) = \frac{1}{|\mathcal{Q}|} \sum_{q_{i}\in \mathcal{Q}}  \big| \big( p(c_{k}, q_{i}) - \max_{c_{j}\in S_{k-1}} p(c_{j}, q_{i}) \big) > 0 \big|
\end{equation}

The overall gain in that case is defined as:
\begin{equation}\label{eq:gain_nbp}
N_{Gain}(c_{k}, S_{k-1}) =  N_{Reward}(c_{k}, S_{k-1}) - (1+\beta) N_{Risk}(c_{k}, S_{k-1})
\end{equation}

\vspace{6pt}
\noindent\textbf{Gain to update the pool of candidate configurations}.

Finally, at step $k$, we select the configuration $c^{*}_{k}$ which maximizes the overall gain according to the following Eq.:
\begin{equation}\label{eq:argmax_gain}
	c^{*}_{k}= \argmax_{c_{k}\in \mathcal{L}\setminus S_{k-1}} \bigg( Gain (c_{k}, S_{k-1}) \bigg)
\end{equation}
Where Gain refers either to Eq. ~\ref{eq:gain_eff} for effectiveness optimization (in that case the gain is the mean improvement of the effectiveness over the training queries)   or Eq. ~\ref{eq:gain_nbp} for the number of improved queries optimization (in that case the gain is the proportion of improved queries).

We then update $S_{k-1}$ as follows:
\begin{equation}
S_{k} = S_{k-1} \cup \{c^{*}_{k} \}
\end{equation}
where $S_{k}$ is the set of $k$  risk-sensitive configurations selected for a set of training queries $\mathcal{Q}$.

The risk-based criteria model we propose is generic enough so that it could be applied to any configuration selection meta-system. 

\begin{figure}[ht]
\centering
\begin{minipage}{.95\linewidth} 
    \normalem 
    \begin{algorithm}[H]
    \SetAlgoLined
	\DontPrintSemicolon 
	\KwIn{$\mathcal{Q}$: The set of training queries, $\mathcal{L}$: A set of system configurations, $\mathbf{M}$: a $\mathcal{L} \times \mathcal{Q}$ matrix containing the effectiveness value of a metric (e.g., AP) for configurations $\mathcal{L}$ over different queries $\mathcal{Q}$, $b$: A baseline configuration, $\beta$: The risk-sensitive parameter, $K$	: The number of configurations to be selected, $E_{Risk}$: the risk function, and $E_{Reward}$: the reward function
	}
	\KwOut{$\mathcal{S}_{K}$: the set of $K$ selected configurations}
	\BlankLine
	$\mathcal{S}_{k} \leftarrow \emptyset $\;

	\tcp*[l]{get the best risk-sensitive configuration given the baseline}
	$c^{*}_{best}$ $\leftarrow$ getBestRSC ($\mathcal{Q}, \mathcal{L}, \mathbf{M}, E_{Risk}, E_{Reward}, \beta, \{b\}$) 

    $\mathcal{L} \leftarrow \mathcal{L} - \{c^{*}_{best}\}$ \tcp*[f]{remove that configuration from the main pool}\;
   
    $\mathcal{S}_{k} \leftarrow \mathcal{S}_{k} \cup \{c^{*}_{best}\}$\;

	$k \leftarrow 1$\;  
	\tcp*[l]{Now the selection uses the current $\mathcal{S}_{k}$ }
	\While{$k < {K}$}{
	$c^{*}_{best} \leftarrow$ getBestRSC ($\mathcal{Q}, \mathcal{L}, \mathbf{M}, E_{Risk}, E_{Reward}, \beta, \mathcal{S}_{k}$)\;
	  $\mathcal{L} \leftarrow \mathcal{L} - \{c^{*}_{best}\}$\;
	  $\mathcal{S}_{k} \leftarrow \mathcal{S}_{k} \cup \{c^{*}_{best}\}$\;
	  $k \leftarrow k + 1$\;
	}
	\Return $\mathcal{S}_{K}$
    \caption{SelectRiskSensitiveConfiguration ($\mathcal{Q}$, $\mathcal{L}$, $b$, $\mathbf{M}$, $\beta$, $K$, $E_{Risk}$, $E_{Reward}$)\; An Algorithm for selecting a pool of configurations using a risk-reward function \label{algo:algorithm1}}
    \end{algorithm}
    \ULforem 
\end{minipage}
\end{figure}

\vspace{5pt}
We present the pseudo-code for selecting $K$ candidate configurations using the risk-reward function in Algorithms~\ref{algo:algorithm1} and~\ref{algo:algorithm2}. To build a set of $K$ candidate configurations, Algorithm~\ref{algo:algorithm1} runs an iterative process that adds a new configuration each time.  
Algorithm~\ref{algo:algorithm2} describes how the best candidate to be added is selected. 

Algorithm~\ref{algo:algorithm1} requires the following input to build the set of $K$ candidate configurations:
\begin{itemize}
    \item The set of training queries $\mathcal{Q}$,
    \item The set of an initial pool of system configurations $\mathcal{L}$,
    \item An effectiveness matrix $\mathbf{M}$ of dimension $\mathcal{L} \times \mathcal{Q}$ containing the effectiveness value of a metric (e.g., AP) for the configurations $\mathcal{L}$ over different queries $\mathcal{Q}$,
    \item The baseline configuration model $b$ (e.g,. BM25),
    \item The risk-sensitive parameter $\beta$,
    \item The number of configurations to be selected $K$,
    \item The risk and reward functions either based on effectiveness using Eqs.~\ref{eq:RISK_eff} ($E_{Risk}$) and~\ref{eq:REWARD_eff} ($E_{Reward}$) or based on the number of queries using Eqs.~\ref{eq:RISK_nbq} ($N_{Risk}$) and~\ref{eq:REWARD_nbq} ($N_{Reward}$).
\end{itemize}

Algorithm~\ref{algo:algorithm2} is slightly  different when $N_{Risk}$ is used while Algorithm~\ref{algo:algorithm1} remains the same.

\begin{figure}[ht]
\centering
\begin{minipage}{.95\linewidth} 
\normalem 
    \begin{algorithm}[H]
	\DontPrintSemicolon
	\KwIn{$\mathcal{S}_{k}$: The set of already selected $k$ configurations, the other parameters are already defined in Algorithm~\ref{algo:algorithm1}}
	\KwOut{$c_{best}$: The configuration with the maximum $gain$}
		\BlankLine	
        $GAIN \leftarrow \{\}$\;		
		\For {$c \in \mathcal{L}$} {
		    $Risk_{\mathcal{Q}} \leftarrow 0$\;
		    $Reward_{\mathcal{Q}} \leftarrow 0$\;
		    \For{$q \in \mathcal{Q}$}{
		        $Risk_{q} \leftarrow$ E$_{Risk}$ ($c$, $q$, $\mathbf{M}$, $\mathcal{S}_{k}$)\;
		        $Reward_{q} \leftarrow$ E$_{Reward}$ ($c$, $q$, $\mathbf{M}$, $\mathcal{S}_{k}$)\;
		        $Risk_{\mathcal{Q}} \leftarrow$ $Risk_{\mathcal{Q}}$ + $Risk_{q}$\;
		        $Reward_{\mathcal{Q}} \leftarrow$ $Reward_{\mathcal{Q}}$ + $Reward_{q}$\;
			}
			$Risk_{c} \leftarrow$ $\frac{Risk_{\mathcal{Q}}}{|\mathcal{Q}|}$\;
			$Reward_{c} \leftarrow$ $\frac{Reward_{\mathcal{Q}}}{|\mathcal{Q}|}$\;
			$Gain_{c} \leftarrow$ $Reward_{c}$ - (1.0 + $\beta$) $\cdot$ $Risk_{c}$\;
			putScorePair ($GAIN$, ($c$, $Gain_{c}$))\;
		}
		$c^{*}_{best} \leftarrow \argmax_{c \in \mathcal{L}} (GAIN)$\;
		\Return $c^{*}_{best}$
    \caption{getBestRSC ($\mathcal{Q}$, $\mathcal{L}$, $\mathbf{M}$, $E_{Risk}$, $E_{Reward}$, $\beta$, $\mathcal{S}_{k}$)\; This Algorithm returns the best configuration using the risk-reward function given the already selected ones~\label{algo:algorithm2}}
    \end{algorithm}
    \ULforem 
\end{minipage}
\end{figure}

\subsection{Selecting the best configuration for a query}
\label{subsec:fit}
\subsubsection{Configuration selection.}
The risk-based criteria we defined aim at optimizing the set of configurations to be used in a per-query configuration selection meta-system. In Deveaud \textit{et al.}~\cite{deveaud2018learningLong} or in Mothe and Ullah~\cite{MotheCIKM2021}, the query-configuration association problem is cast as a learning to rank problem where L2R algorithms are applied to decide which of the configuration from the candidate  set should be used. He and Ounis~\cite{he2004query}  rather considered a classification model where queries are clustered and a configuration is associated to each cluster.

{The rationale has been that L2R worked well in various contexts such as learning to rank documents, learning to rank terms to add to a query in query expansion. We thus developed a model to adapt the principle of L2R documents to L2R configurations so that the L2R model learns to rank the possible configurations in order to select the best one~\cite{deveaud2018learningLong,MotheCIKM2021}. In traditional L2R documents, the model's goal is to learn a global ranking function for all the queries while in L2R configurations, the model aims at determining a query-specific retrieval process. Intuitively that sounded and worked. To implement this, we needed examples, they consisted in query features and configuration features plus a label which was the effectiveness on the query-configuration pair. }

In this L2R configuration pipeline, after training, when ranking multiple configurations and choosing the best one for treating unseen or test query, the query features are {obviously} the same across the testing examples and the configuration features (component used, hyperparameters used) are different. We observe that the configuration features are not discriminatory enough to differentiate the candidate configurations. Therefore, the L2R configuration model may not consistently predict the best configuration to treat a query. To solve this issue, here, we choose a much simpler method. For each training query, we select the best configuration from the pool of $K$ risk-sensitive configurations and assign it to the training query. After the model is trained, for unseen queries,  we search the best-matched training query based on the Cosine  similarity between the testing query and the training query in terms of query features. We use the configuration associated with the best-matched training query for each testing query. This is similar to first-nearest-neighbor method.

The hypothesis behind this approach is that although queries are indeed diverse, some search queries share common retrieval characteristics that could help in deciding how the query should be processed. If we could represent the queries with appropriate features that help in making the link between the query and the best query processing to be applied, we could then rely on these features for unseen queries. The approach is related in some ways to query performance prediction, but here, we do not predict the performance but rather the query processing that should be applied. It is also linked to multi-task learning where  one query could be considered as one task (finding the best configuration) and the similar tasks could be tackled by similar treatments (similar configurations). We keep the analysis on these points for future study.

\begin{figure}[ht]
\centering
\begin{minipage}{.95\linewidth} 
\normalem 
    \begin{algorithm}[H]
	\DontPrintSemicolon
	\KwIn{$test_{f}$: The query features for the test query, other parameters are already defined in Algorithm~\ref{algo:algorithm1}}
	\KwOut{$c^{*}_{best}$: The best-match configuration for the given test query}
		\BlankLine	
		$\mathcal{I} \leftarrow \{\}$\;
		\For {$q \in \mathcal{Q}$} {
		    $bc \leftarrow$ getBestSC ($q, S_{k}, M$)\;
		    putScorePair ($\mathcal{I}, (q, bc)$)\;
		}
		$CS \leftarrow \{\}$\;
		\For{$q \in \mathcal{Q}$}{
		  $q_{f} \leftarrow$ getQueryFeatures ($q$)\;
		  $score \leftarrow$ computeCosine ($q_{f}, test_{f}$)\; 
		  putScorePair ($CS, (q, score)$)\;
		}
		$q^{*}_{best} \leftarrow \argmax_{q \in \mathcal{Q}} (CS)$\;
        $c^{*}_{best} \leftarrow$ getScore ($\mathcal{I}, (q^{*}_{best})$)\;
        \Return $c^{*}_{best}$
    \caption{bestMatchConfiguration($\mathcal{Q}$, $\mathcal{S}_{k}$, $\mathcal{M}$, $test_{f}$)\; This Algorithm returns the best match configuration~\label{algo:algorithm3}}
    \end{algorithm}
    \ULforem 
\end{minipage}
\end{figure}

In concrete terms, the idea of this simple yet effective approach is quite straightforward. We present the process of selecting the best configuration for a given test query in Algorithm~\ref{algo:algorithm3}. Given the set of training queries $\mathcal{Q}$, the set of risk-reward based selected configurations $\mathcal{S}_{k}$, the effectiveness matrix $\mathcal{M}$ of training queries $\mathcal{Q}$ by configurations $\mathcal{S}_{k}$, and the query feature vector $test_{f}$ of testing query, we first develop an index, $\mathcal{I}$ for keeping the best configuration for each training query from Steps 2 to 4. That is, for each training query $q \in \mathcal{Q}$, we search the best configuration $bc$ from $\mathcal{S}_{k}$ and assign the $bc$ in $\mathcal{I}$ by associating it with the training query $q$. Second, From Steps 7 to 10, we compute the Cosine  similarity score for the set of training queries $\mathcal{Q}$ with the given test query using query features-based representations. Third, in Step 12 we find the training query which gets the maximum similarity with the testing query. Finally, in Step 13, we look up the respective configuration from the query-configuration index $\mathcal{I}$ and recommend it for treating the underlying test query.

\vspace{6pt}
\noindent\textbf{Query representation.}
With regard to \textit{query features}, our framework is generic enough so that we could use any kind. {In the literature, many different query features have been used for different tasks.}  Query features have been defined for L2R documents~\cite{macdonald2013learning,cao2007learning,lin2017learning} or query difficulty prediction~\cite{Mothe2005,guo2010predicting,shtok2012predicting}. We adapt the query features usually used for L2R documents~\cite{cao2007learning,lin2017learning}. We choose to use LETOR features that were successfully used in both L2R documents and query performance prediction application~\cite{Chifu2018Long}. In L2R documents, LETOR features indicate how relevant or important the document is with respect to the query~\cite{qin2010letor}. Examples of such features are the BM25 score, the term frequency of the query terms, etc. Each of these features is associated with a query-document pair.

Here, instead of features associated with query-document pairs, we need features associated with query-configuration pairs since the training is to decide which configuration better fits a query. LETOR features are query-document scores; they thus could be calculated for each query and each configuration $c_{i}$. It would be, however, quite costly. Thus, we consider a unique fair configuration (in our experiments, we chose $c_{BM25}$ as this reference configuration -BM25 without query expansion) to calculate the LETOR query features. This option is a reasonable trade-off between computing costs and accurate query representation.

Since the LETOR functions $\phi_L$ 
are associated with query-document pairs rather than query-configuration pairs, we need to aggregate them over the documents that have been retrieved for a query. Let $d_.^i = (d^{i}_{1}, d^{i}_{2},\cdots, d^{i}_{n} )$ be the $n$ top retrieved documents for the given query $q_{i}$ based on the reference configuration. We can estimate a set of scores $W_{i}  = \{w^i_L\}$, where $w^i_L= \mho(\phi_L(d^i_., q_{i}))$ with $\mho$ a set of aggregation functions and $\phi_L$ a set of LETOR scoring functions. A similar principle was used  for query performance prediction~\cite{Balasubramanian2010,Chifu2018Long}. With regard to $\mho$, we can use any kind of statistical aggregation functions.

In the next section, we present the evaluation framework of our risk sensitive model.
\section{Evaluation}
\label{sec:eval}
\subsection{Objectives of the experiments}
We developed a series of experiments that aim at evaluating several aspects of the proposed risk-based functions 
as follows:

\begin{itemize}
    \item {After the generation of the configurations, the phase 2 of the model training (see Figure~\ref{fig:Overview}), the model selects the $k$ configurations using our risk functions (phase 2), then phase 3 is selective query processing that decides which of the $k$ configurations should treat a given query. We need to evaluate this framework as much as possible;} 
    \item The results may be collection-dependent; it is thus important to evaluate the results on various collections. We chose three standard TREC reference collections for ad hoc (see Section~\ref{subsec:EriskAdhoc} and Table~\ref{tab:GOV2eff_APR_20_20} and \ref{tab:trec78WT10Geff_APR_20_20}). These different collections are used across the experiments. 
   
    \item The results may be task-dependent; evaluating the model on different tasks is important. We consider both ad hoc search and diversity-oriented search. Diversity task is covered in Section~\ref{subsec:EriskDiversity} and in Table~\ref{tab:diversity09b12b_APR_20_20} for two different collections. 
    \item The candidate configuration selection relies on an iterative process that adds a new configuration at each iteration; the optimal number of configurations is worth studying for both tasks (see Section~\ref{subsec:NbConfig} and Figure~\ref{fig:kvariants});
      
    \item We developed two risk-sensitive functions to select the candidate configurations, one is oriented toward overall effectiveness, $E_{Risk}$ and the other is oriented toward the number of queries {for which effectiveness is improved}, $N_{Risk}$. These two functions need to be compared (see Section~\ref{subsec:2func} and Table~\ref{tab:Nrisk});

    \item Most of the ad hoc or diversity collections have deep relevance judgments. We also evaluate the result on an additional shallow relevance judgment collection, MS MARCO (see Section~\ref{Sub:MS-MARCO} and Tables~\ref{tab:MS-MARCO-Dev_20_20} and~\ref{tab:MSMARCO-dev_rbp_10}).
    
    \item {Some insights on the results are discussed in Section~\ref{sub:deep}}.
    
\end{itemize}

\subsection{{Configuration settings}}
\label{sub:EvalConfigurations}

We consider two search components: the retrieval model used (e.g., LM, BM25, etc.), the query expansion model (e.g., Bo2), and three hyperparameters from the query expansion component: the number of documents, the number of terms, and the minimum number of documents in which the terms should occur. We made these choices based on our previous studies that show that the retrieval and query expansion module are the most important, specifically for difficult queries, the hyperparamerters of query expansion are of help~\cite{compaore2011mining}. The possible values we considered for these five features are stated in Table~\ref{tab:paravalues}. Alternatively, we could have considered the retrieval component hyper parameters, however, we keep this analysis and related experimental part for future work.

We used Terrier that implements these retrieval and query expansion models. For retrieval models, we used Terrier default hyperparameter values.

\begin{table}[ht!]
\caption{Configuration features (components and hyperparameters) and their values.}{
\begin{tabular}{@{}l@{\hspace*{0.1cm}}|@{\hspace*{0.1cm}} l@{}}
Parameters & Values\\
\hline
\multirow{4}{2.5cm}{Retrieval model}
& {BB2, BM25, DFRee, DirichletLM, HiemstraLM,} \\
& {InB2, InL2, JsKLs, PL2, DFI0, XSqrAM, DLH13,} \\
& {DLH, DPH, IFB2, TFIDF, InexpB2, DFRBM25,}\\
& {LGD, LemurTFIDF, InexpC2}\\
\hline
\multirow{2}{2.5cm}{QE model}
& {No, KL, Bo1, Bo2, KLCorrect, Information,}\\
& {KLComplete}\\
\hline
\multirow{1}{2.5cm}{\# of Exp. doc.}
&  2, 5, 10, 20, 50, 100\\
\multirow{1}{2.5cm}{\# of Exp. terms}
&  2, 5, 10, 15, 20\\
\multirow{1}{2.5cm}{Min. \# of doc.}
& 2, 5, 10, 20, 50\\
\end{tabular}\label{tab:paravalues}}
\end{table}

Combining all these search parameter values from Table~\ref{tab:paravalues} makes more than $20,000$ possible configurations. These configurations are the ones we mentioned in the first phase of the process as depicted in Figure~\ref{fig:Overview}.

\subsection{{Query features and labels for training}}
\label{subsec:Query}
\vspace{6pt}
\noindent\textbf{Query features.} {We need features to represent queries.}
Given a query, we extract the LETOR features (LETOR scoring functions $\phi$) that are implemented in the Terrier FAT framework\footnote{\url{http://www.terrier.org/docs/v4.0/javadoc/index.html?org/terrier/matching/models/WeightingModel.html} provides the details of the features and their implementation}~\cite{macdonald2013learning}. The Terrier FAT framework uses DAAT~\cite{turtle1995query} retrieval strategy to identify the initial sample of $n$ documents and keeps all the posting of these documents in memory. This strategy allows fast computation of any scoring function (i.e., weighting function) without resorting the expensive inverted index in real-time.

We extract the query-document features using Terrier's weighting model\footnote{\url{http://terrier.org/docs/v5.2/javadoc/org/terrier/matching/models/WeightingModel.html}}) (See Table~\ref{tab:QueryDocFeatures}).

\begin{table}[ht!]
\caption{Query-document features from Terrier's weighting models}{
\begin{tabular}{ l @{\hspace*{0.5cm}} l @{\hspace*{0.5cm}} l @{\hspace*{0.5cm}}  r@{}}
 WMODEL:Tf & WMODEL:TF\_IDF & WMODEL:LemurTF\_IDF \\
 WMODEL:BM25 & WMODEL:Js\_KLs & WMODEL:In\_expC2 \\
 WMODEL:InB2 & WMODEL:DLH &  WMODEL:ML2 \\
 WMODEL:BB2 &   WMODEL:DFIC & WMODEL:IFB2 \\
 WMODEL:InL2 & WMODEL:PL2 & WMODEL:LGD \\
   WMODEL:MDL2 & WMODEL:DirichletLM & WMODEL:DFRee \\
   WMODEL:Hiemstra\_LM
\end{tabular}\label{tab:QueryDocFeatures}}
\end{table}

LETOR features are extracted for the initial documents retrieved based on the reference configuration $c_{BM25}$~\cite{macdonald2013learning}. 
We aggregate these features over the documents for a given query as to obtain LETOR features at a document set level~\cite{Chifu2018Long,Balasubramanian2010}. We use the mean, standard deviation, and maximum summary functions to obtain the feature vectors that represent each query-configuration pair. We chose these aggregation functions as Chifu \textit{et al.} showed that they are complementary for query performance prediction~\cite{Chifu2018Long}. 

\vspace{6pt}
\noindent\textbf{Labels for the training examples and evaluation measures.}
For the training queries, the label is the effectiveness of the configuration when treating the query. We keep both the configuration and the associate effectiveness for each query and each configuration. We performed a series of experiments in which the type of label, that is to say, the effectiveness measure we use, changes. We considered successively different types of labels, either ad hoc-oriented or diversity-oriented effectiveness measures (see Subsection~\ref{subsec:Evaluation_measures}).

\subsection{Data collection}
We considered six standard TREC collections: TREC78, WT10G, GOV2, MS MARCO from the ad hoc task and Clueweb09, and Clueweb12 from the diversity task.

For TREC78, there are approximately 500K newspaper articles. The WT10G collection is composed of approximately 1.6 million Web/Blog page documents. GOV2 collection includes 25 million web pages, a crawl of .gov domain. MS MARCO collection includes 3.2 million documents. Clueweb09B and Clueweb12B contain 50 and 52 million web pages. 

The  TREC test collections include topics. The ``standard" format of a TREC topic statement comprises a topic ID, a title, a description, and a narrative. The title contains two or three words on average that represent the keywords a user could have used to issue a query to a search engine. We use the title part only in the evaluation.

\begin{table}[ht!]
\caption{Statistics of the collections used.}{
\begin{tabular}{@{} l@{\hspace*{0.5cm}} l @{\hspace*{0.5cm}} c @{\hspace*{0.5cm}} c @{\hspace*{0.5cm}} r@{\hspace*{0.5cm}} r@{}}
& Collection & \#Docs & Queries (title only) & Avg. Rel. & Avg. Irrel.\\
\hline
\multirow{3}{1.2cm}{Ad hoc}
& TREC78 & 528K & 100 (351 -- 450) & 92.02 & 1577.73\\
& WT10G & 1.692M & 100 (451 -- 550) & 59.80 & 1344.90\\
& GOV2 & 25M & 150 (701 -- 850) & 179.45 & 722.90\\
& {MS MARCO} & {3.2M} & {5,193 (Dev set)} & 1.00 & 0.00\\
\hline
\multirow{2}{*}{Diversity}
& Clueweb09B & 50M & 200 (001 -- 200) & 166.26 & 633.45\\
& Clueweb12B & 52M & 100 (201 -- 300) & 197.50 & 689.04\\
\end{tabular}\label{tab:collection}}
\end{table} 

Table~\ref{tab:collection}  summarizes a few features about these collections. There are 100 topics in TREC78 (merging topics from TREC7 and TREC8), 100 topics in WT10G, 150 topics in GOV2, 5193 topics in MS MARCO (Development set), 200 topics in Clueweb09B (merging the 2009 to 2012 web track queries), and 100 topics in Clueweb12B (merging the 2013 and 2014 queries). 

Finally, the collections provide \textit{qrels} (i.e., judged documents with eventually graded relevance for each topic). 
Table~\ref{tab:collection} also reports the average number of judged relevant (Avg. Rel.) and irrelevant (Avg. Irrel.) documents for each of the collections. Except for MS MARCO, each collection has deep relevance judgments. MS MARCO has a shallow relevance judgment and has only one relevant document per query. 

\subsection{Evaluation measures}
\label{subsec:Evaluation_measures}
We used the common evaluation measures to estimate the retrieval effectiveness for a query with a configuration.

\begin{itemize}
\item Precision at the cut-off 10 documents (P@10), Average Precision (AP)~\cite{carterette2011overview,sakai2007reliability}, normalized Discounted Cumulative Gain (nDCG)~\cite{jarvelin2017ir}, and Ranked Biased Precision (RBP)~\cite{rbp-moffat-zobel} documents are used for ad hoc search evaluation of the queries from TREC78, WT10G, GOV2, and MS MARCO collections (in the latter AP is replaced by RR). We use evaluation program \textit{trec\_eval}\footnote{http://trec.nist.gov/trec\_eval/} to estimate AP, nDCG@10, and P@10 metrics and 
\textit{rbp\_eval}\footnote{https://github.com/jsc/rbp\_eval} to compute the RBP metric with persistence levels 0.50, 0.80, and 0.95, respectively. 
The P@k and AP are defined based on binary relevance, while nDCG@k is based on graded relevance. The RBP metric is defined based on either binary or graded relevance. Unlike other metrics, RBP metric includes a user persistence parameter which can be used to observe users' browsing behavior in the ranked list. Moreover, the RBP metric helps us to explore the limits of residual error, considering the unjudged documents in the ranked list as judged and relevant. Thus, the analysis of the effect of shallow vs. deep relevance judgments could be feasible using RBP metric.

\item Intent-aware expected reciprocal rank at the cut off 20 (ERR-IA@20) (Official metric)~\cite{chapelle2009expected}, $\alpha$-nDCG@20 where $\alpha$ is set to 0.5, and novelty \& rank-biased precision (NRBP)~\cite{clarke2008novelty} are used for diversity-oriented search evaluation on Clueweb09B and Clueweb12B collections. We use evaluation program \textit{ndeval}\footnote{https://trec.nist.gov/data/web/11/ndeval.c} to calculate the effectiveness measures for diversity metrics.
\end{itemize}

\subsection{Cross-validation} 
Learning is based on the training queries, then the trained meta-search model is applied to the test queries. We used two-fold cross validation to ensure independent sets and keep enough queries in both the training and test sets~\cite{wong2017dependency}.
We proceeded as follows: half of the queries, let us call this query set $Q_A$, are used for training while the other half, $Q_{\overline A}$, is used for testing. Then, in a reverse sense, $Q_{\overline A}$ is used for training while $Q_A$ is used for testing. We made $3$ draws to randomly split queries into $Q_A$ and $Q_{\overline A}$ and averaged the results. To prepare the folds for three trials, we randomly shuffled the queries using R's sample function (random seed=42) and divide the query set into two subsets for each trial. We report the mean and standard deviation across the three trials and two directions. Thus, each mean and standard deviation is over a total of six measurements. The same splits are used whatever the method that needs training.

For statistical significance analysis, we use two-tailed paired t-test with Bonferroni correction (p-value $<$ 0.05).

\subsection{Baselines and related work methods}
\label{subsec:baseline}

\vspace{6pt}
\textbf{Non selective baselines.} 
We consider the following baseline methods for which the same process is applied what ever the query is (non-selective query processing methods thus) :

\begin{itemize}
    \item BM25 is obtained using Terrier BM25 with default parameters b=0.75 and k=1.2 (default hyperparameter values)~\cite{robertson2009probabilistic} on the entire set of the queries;
    \item L2R-D SVM$^r$ is the usual learning to rank documents where the initial ranking is obtained using BM25. L2R is based on SVM-rank for which we obtained the best results compared to other L2R algorithms (we tried Random Forest, SVM-rank, $\lambda$-MART and LISTNet). L2R ranks needs positive and negative examples. To select positive and negative examples, we consider the relevance judgment of each collection. Given a query the documents with an effectiveness label greater than 0  are considered positive examples, while documents labeled as less than or equal to 0 are considered negative examples. If there is no document labeled with a negative grade, we retrieve 1,000 ranked documents as baseline retrieval for a query using BM25 model and consider the same number of positive documents from bottom-ranked one of baseline retrieval as the negative examples;
    \item GS is Grid Search that learns the best value for each parameter to maximize the effectiveness for a set of queries~\cite{taylor2006optimisation};
    \item Best trained is the best configuration among the entire configuration set, the one that maximized the effectiveness measure considered: it is chosen for $Q_{A}$ and applied to $Q_{\overline{A}}$ and vice versa;
    \item CombSum: It implements data fusion as the combinations of system document lists; we consider the CombSum~\cite{fox1994combination} on the two first configurations selected by $E_{Risk}$. \end{itemize}

Among all these baselines, BM25 is the only one that does not need training but rather is applied on the entire query set. Thus, BM25 is deterministic. The other  baselines are reported considering 2-fold cross-validation for three random trials. The same three splits are used whatever the method is. We report the average over the three trials and the two test folds as well as the standard deviation.

\vspace{6pt}
\textbf{Selective approaches.}
 
In the evaluation section, in addition to our selective query processing strategy (ERisk-Cosine), we report :
\begin{itemize}
    
    \item Trained SQE considers two configurations: one with query expansion and one without query expansion. We took the best configuration ("Best conf.") and its query expansion/non query expansion counterpart. In cases where the best configuration did not include query expansion, we chose the best configuration with query expansion from the pool as its counterpart. Then we trained the meta-system for selective query expansion.
    \item Random20-RF: L2R configuration selection as presented in Deveaud \textit{et al.}~\cite{deveaud2018learningLong}. Like in L2R documents, for L2R configuration, the ranking model $R(q_{i}, \mathcal{S})$ can learn from the training data by minimizing the loss function $\mathcal{L}(r; \mathbf{f}, \mathcal{S})$. The results with various  L2R models (Linear Regression, Random Forest (RF), SVM-Rank, and LambdaMart~\cite{lambdamart}) are not significantly different. We report RF only. Here the 20 candidate configurations are selected randomly.
    
    \item ERisk-RF: Like Random20-RF but here the configuration selection is using the $E_{Risk}$ function, following the framework of our previous work in CIKM~\cite{MotheCIKM2021}.
\end{itemize}

The ERisk-Cosine is similar to ERisk-RF but here, the selective query processing step does not rely on a heavy training but just on the query feature similarities; which is an extension of this current paper. 
All these methods are based on a training/testing principle. We use exactly the same three random splits as for the baselines that need training; we report the average and standard deviation.

\vspace{6pt}
\textbf{Oracles.}
We also report the following Oracle methods:
\begin{itemize}  
    \item Best conf.: the best configuration in the pool of configurations; the one that maximizes the considered evaluation measure (the configuration may be different according to the measure);
    \item Oracle: for each of the queries, this is the best configuration among the $20,000$ possible configurations;
    \item Oracle20ERisk: for each of the queries, this is the best configuration among the $20$ possible configurations selected using the corresponding Risk function, and
    \item Oracle20Random: for each of the queries, this is the best configuration among the $20$ possible configurations selected randomly.
\end{itemize}
Oracle methods are not deterministic: it is impossible to select them automatically. Rather, Oracles would need a perfect system that chooses the most appropriate configuration per query that in practice can not be obtained automatically. We report the result for the first random split for Oracles.

\section{Results and discussion}
\label{sec:res}

In the first series of experiments, we evaluate the  $E_{Risk}$ function with a fixed number of configuration candidates on ad hoc retrieval (Section \ref{subsec:EriskAdhoc}) and on the diversity task (Section~\ref{subsec:EriskDiversity}). We analyze  the impact of $k$, the number of configuration candidates (Section~\ref{subsec:NbConfig}) and compare $E_{Risk}$ and $N_{Risk}$  (Section \ref{subsec:2func}) on both tasks. We study the specific case of sparse feedback with the MS MARCO data set (Section~\ref{Sub:MS-MARCO}). Finally, we provide some insights into our results (Section \ref{sub:deep}).

\subsection{$E_{Risk}$ function for ad hoc search}
\label{subsec:EriskAdhoc}

Our framework consists of two  selective stages the first being to select the most appropriate $k$ candidate configurations from the entire configuration set and then to choose the best configuration for 
each query. 

To evaluate the impact of the $E_{Risk}$ function, we compare its results against where  configurations are randomly selected as in Deveaud \textit{et al.}~\cite{deveaud2018learningLong}. We consider a fixed number of configurations. Following initial experiments we set this number to 20 (see also Section~\ref{subsec:NbConfig}). We thus consider either 20 randomly\footnote{We use \url{https://docs.python.org/3/library/random.html\#random.sample function}} selected configurations (Random20-RF row) or 20 configurations chosen using $E_{Risk}$. These $20$ configurations are then used in phase 3 by either the L2R configuration-based model\footnote{The results slightly differ from the ones in Mothe and Ullah~\cite{MotheCIKM2021} where we consider 20 configurations in phase 2 but only 2 configurations in phase 3.} (ERisk-RF) or the Cosine -based model (ERisk-Cosine).

\begin{table}[!ht]
\caption{Our Risk-based models significantly outperform any other non-oracle baseline on ad hoc GOV2. The results are averaged based on 3 draws and  two-fold testing. The left column of results reports absolute values and  standard deviations in square brackets. For our methods (ERisk-xx), $\vartriangle$ (resp. ${\uparrow}$) indicates statistically significant improvement over the L2R documents (resp. grid search (GS)) according to a paired t-test ($p < 0.05$). On the right column we indicate the percentage increase or decrease when compared to L2R documents and grid search (GS) in parenthesis. The model selects and uses $k=$20 candidate configurations in both phase 2 and 3. Although the heading of the metrics is AP, nDCG@10, and P@10, the scores in the Table are the mean of these metrics over queries (e.g., MAP for AP).}{
\centering
\begin{tabular}{@{}l@{\hspace*{0.2cm}} l@{\hspace*{0.1cm}}|@{\hspace*{0.1cm}}l@{\hspace*{0.1cm}} l@{\hspace*{0.1cm}} l@{\hspace*{0.1cm}} c@{}| @{\hspace*{0.1cm}} l@{\hspace*{0.1cm}} l@{\hspace*{0.1cm}} l@{}}

\multicolumn{2}{l}{$E_{Risk}$} & \multicolumn{6}{c}{GOV2}\\
\hline
& & \multicolumn{3}{c}{Absolute values} && \multicolumn{3}{c}{Percentage (\%) }\\
\cline{3-9}
& Methods & AP & nDCG@10 & P@10 && AP & nDCG@10 & P@10 \\
\hline
\multirow{5}{*}{\rotatebox[origin=c]{270}{\textbf{Baselines}}}
& BM25 & .27$^{~~}$$^{~~}$ & .46$^{~~}$$^{~~}$ & .54$^{~~}$$^{~~}$ && -04\% (-23\%) & -06\% (-12\%) & -05\% (-13\%)\\

& L2R-D SVM$^r$ & .28$^{~~}$[.001] & .49$^{~~}$[.002] & .57$^{~~}$[.003] && na (-20\%) & na (-06\%) & na (-08\%)\\

& GS & .35$^{~~}$[.005] & .52$^{~~}$[.003] & .62$^{~~}$[.008] && +25\% (na) & +06\% (na) & +09\% (na)\\

& Best trained & .35$^{~~}$[.005]$^{~~}$ & .49$^{~~}$[.012]$^{~~}$ & .59$^{~~}$[.010]$^{~~}$ && +25\%(+00\%) & +00\% (-06\%) & +04\% (-05\%)\\

& CombSUM & .36 [.003] & .54 [.005] & .65 [.008] && +29\% (+03\%) & +10\% (+04\%) & +14\% (+05\%)\\
\hline

\multirow{4}{*}{\rotatebox[origin=c]{270}{\textbf{SQP}}}
& Trained SQE & .35 [.005]$^{~~}$ & .52 [.003]$^{~~}$ & .63 [.007]$^{~~}$ && +25\% (+00\%) & +06\% (+00\%) & +11\% (+02\%)\\

& Random20-RF 
& .33$^{\vartriangle}$$^{\downarrow}$ [.003] & .48$^{~~}$$^{\downarrow}$ [.007] & .58$^{~~}$$^{~~}$ [.009] && +17\% (-05\%) & -02\% (-08\%) & +02\% (-06\%)\\

& ERisk-RF & .39$^{\vartriangle}$$^{\uparrow}$ [.002] & .61$^{\vartriangle}$$^{\uparrow}$ [.010] & .73$^{\vartriangle}$$^{\uparrow}$ [.015] && +39\% (+11\%) & +24\% (+17\%) & +28\% (+17\%) \\

&\textbf{ERisk-Cosine} & \textbf{.40}$^{\vartriangle}$$^{\uparrow}$ [.003] & \textbf{.62}$^{\vartriangle}$$^{\uparrow}$ [.003] & \textbf{.76}$^{\vartriangle}$$^{\uparrow}$ [.019] &&+43\%(+14\%)&+27\% (+19\%) & +33\% (+23\%)\\
\hline
\multirow{4}{*}{\rotatebox[origin=c]{270}{\textbf{Oracles}}}
& Best conf. & .36 & .52 & .63 &&  +29\% (+03\%) & +06\% (+00\%) & +10\% (+02\%) \\
& Oracle20ERisk & .42$^{~~}$ & .68$^{~~}$ & .80$^{~~}$ && +50\% (+20\%) & +39\% (+31\%) & +40\% (+29\%)\\
& Oracle20Random 
& .41 $^{~~}$ & .68 $^{~~}$ & .79 $^{~~}$ && +46\% (+17\%) & +39\% (+31\%) & +39\% (+27\%)\\
& Oracle & .50$^{~~}$ & .85$^{~~}$ & .94$^{~~}$ && +79\% (+43\%) & +73\% (+63\%) & +65\% (+52\%)\\
\end{tabular}\label{tab:GOV2eff_APR_20_20}}
\end{table}

Tables \ref{tab:GOV2eff_APR_20_20} and \ref{tab:trec78WT10Geff_APR_20_20} report the results on the three ad hoc collections for $E_{Risk}$. Table~\ref{tab:GOV2eff_APR_20_20} reports the results for the $E_{Risk}$ function on the GOV2 collection; this is the largest 
collection of the three ad hoc collections we considered. Horizontally, the first block reports the  baselines we described in Section~\ref{subsec:baseline}, the second block is the variants of the risk-based and selective query processing approaches 
and finally the last block is the Oracles. The left column of results in the table reports the absolute values of the measures when averaged over three different random splits for 2-fold cross validation. We also report the standard deviation in square brackets. On the right column, we report the percentage increase/decrease when compared to, on the one hand L2R documents and on the other hand grid search (in parenthesis). 

From these results, we can observe that $E_{Risk}$ models (the two last rows from the 2nd horizontal block, the first one as presented at CIKM and the second being specific to this extended version)  outperform all the baselines (1st horizontal block) as well as the other selective approaches when considering randomly selected configurations. The differences are statistically significant when compared to both grid search (${\uparrow}$) and L2R document ($\vartriangle$) baselines. 

Compared to previous selective query processing strategies, the very impact of the second phase of the process, the risk-based selection, can be evaluated by comparing Random20-RF and both ERisk-RF, where the third phase is identical for both.  
The impact of the third phase of the process 
can be evaluated when comparing ERisk-RF and ERisk-Cosine. Finally, the overall process can be evaluated when comparing ERisk-Cosine and L2R-D SVM; both make use of LETOR features in a training process. The overall process can also be compared to a single optimal configuration (e.g. GS) and to data fusion (CombSUM).

From the results in Table~\ref{tab:GOV2eff_APR_20_20}, we can conclude that the component of the process with greatest impact is  the configuration selection part that we developed in this research. Within the selective query processing (SQP) block, when we compare the randomly selected configurations and the $E_{Risk}$ selection, we can observe that $E_{Risk}$ models outperform (statistically significantly) the Random one. We can also observe that the results are robust according to the standard deviation across the three random train/test splits. 

Among  more than $20,000$ different configurations, the configuration that maximizes P@10 (Best conf. in Table~\ref{tab:GOV2eff_APR_20_20}) obtains 0.63 (InB2 retrieval model with expansion model KLCorrect with 5 expansion documents, 20 expansion terms, and 2 minimum documents that those terms should occur in); the same configuration also maximizes nDCG@10 and obtains 0.52. 
The one that maximizes AP obtains 0.36 (DFRee model with expansion model Bo1 with 20 expansion documents 20, 20 expansion terms, and 5 minimum documents). The best configuration from among the $20,000$ is also the best configuration from among the 20  $E_{Risk}$ functions. By definition, the best configuration for a given measure cannot be surpassed by any other single configuration. It is not possible however to decide in advance what will be the best configuration for a given collection as this is something only an Oracle can do.

The best trained configuration (second to last row in the baseline block) uses exactly the same train/test splits as our models and selects a single configuration based on the training queries. This selected configuration is then used for all the test queries. It is thus  a fair baseline since, as our methods, it can be trained. Best trained model obtains P@10 of 0.59. It is important to mention that none of our initial configurations use document re-ranking. This is what L2R-D (L2R documents) does (2nd line in the baseline block). L2R-D surpasses BM25 on which it is based for the initial document ranking but is not the best trained on P@10 or AP. Some other single configurations, including the best, surpass L2R-D but this best configuration cannot be decided prior to evaluation. The grid search (GS) is able to find a configuration close to the best configuration (P@10 is 0.62). The same comments hold also for AP and nDCG@10. GS is the best among the baselines that use a single configuration, then best trained and L2R-D are relatively close one 
another.  Both require training and both use a single configuration for all the test queries. CombSUM  uses several configurations for which the resulting document lists are fused, surpasses the  baselines that use a single configuration. CombSUM is very close to the Train SQE method and to the best configuration oracle. All the selective query processing methods have, however, the further advantage of single query processing for each query once decided, while CombSUM needs to evaluate several and thus CombSUM is less efficient. The $E_{Risk}$ based models are also much more effective than CombSUM.

Regarding the variants of our models that differ according to the third phase, either Cosine ($E_{Risk}-Cosine$) or L2R using random forest ($E_{Risk}-RF$), we can see that the simplest method (Cosine) is also the most effective. The reason may be because of the quality of the representation: in learning to rank, the training features consist of the LETOR features plus the configuration features where configuration features are the ones presented in Table~\ref{tab:paravalues} transformed into numerical values. This representation is less effective than in the second case where we compare queries with LETOR features only.
We used three trials for randomly splitting training and test queries. As we mentioned earlier, the table reports the average values and standard deviations. Most of the standard deviations are in of the order of $10^3$, which means the method is robust. This also holds  for baseline results. 

Finally, Oracles indicate the maximum effectiveness that we could reach with a perfect match between a query and the best configuration for that query,  whatever the query is. Although our models can still be improved to reach these maxima, we can observe that the training works pretty well and our models are at the midway point when considering the best configuration of the pool on the one hand (e.g., P@10 0.59) and the oracle on the other  (e.g., P@10 0.80). Oracle20ERisk is when the perfect query-configuration match is based on the 20 configurations that are selected by the $E_{Risk}$ function, Oracle20Random is when this perfect match is based on 20 randomly selected configurations while Oracle is when all  $20,000$ configurations are used and there is no risk function applied. Comparing Oracle20Erisk and Oracle we  see that although Oracle is better than Oracle20ERisk, the decrease in the number of configurations by a factor of 1,000 (from $20,000$ to 20) does not drastically affect  performance. This shows that the candidate selection phase with $E_{Risk}$ is effective.

The right column of Table~\ref{tab:GOV2eff_APR_20_20} reports the percentage of decrease/increase compared to L2R-D and GS. We can see that using different configurations rather than a single one in an adaptive way leads to +11 to +17 \% for ERisk-RF (+14 to +22 \% for ERisk-Cosine) of effectiveness when compared to using a single optimized configuration (grid search). We think percentages are also worth reporting to show the impact of our model.

More surprising are the results of Oracle20Random compared to Oracle20Erisk in Table~\ref{tab:GOV2eff_APR_20_20} (e.g., P@10 is 0.79 and 0.80). These results imply that even randomly chosen configurations can lead to high effectiveness if they are appropriately associated with the queries to process. There are two possible reasons for this. First GOV2 is indeed an ``easy" collection where a majority of configurations work well on  queries. The second is  that some configurations may be effective by chance but it is likely that training will not generalize well on these configurations.

We looked at the number of queries that were improved by our method using $E_{Risk}$. For example, when considering $ERisk-RF$ and P@10 on GOV2, there are 90 queries that are improved compared to GS; on TREC78 and WT10G, there are 73 queries that are improved. 

\vspace{6pt}Table~\ref{tab:trec78WT10Geff_APR_20_20} reports the results for both TREC78 and WT10G collections, still for $E_{Risk}$ function and with $20$ configurations.
\begin{table}[!ht]
\caption{The good performance of our Risk-based models generalized well on ad hoc search $k=20$ configurations on the TREC78 collection (left column) and WT10G collection (right column) for $E_{Risk}$. We use the same notation as used in Table~\ref{tab:GOV2eff_APR_20_20}.
}{
\centering
\begin{tabular}{@{}l@{\hspace*{0.2cm}} l@{\hspace*{0.1cm}}|@{\hspace*{0.05cm}}l@{\hspace*{0.1cm}} l@{\hspace*{0.1cm}} l@{\hspace*{0.1cm}} c@{}| @{\hspace*{0.05cm}} l@{\hspace*{0.1cm}} l@{\hspace*{0.1cm}} l@{}}
& $E_{Risk}$ & \multicolumn{3}{c}{TREC78} && \multicolumn{3}{c}{WT10G}\\
\hline
& Methods & AP & nDCG@10 & P@10 && AP & nDCG@10 & P@10\\
\hline
\multirow{4}{*}{\rotatebox[origin=c]{270}{\textbf{Baselines}}}
& BM25 & .21$^{~~}$$^{~~~}$ & .47$^{~~~}$$^{~~~}$ & .45$^{~~~}$$^{~~~}$ && .22$^{~~}$$^{~~}$ & .39$^{~~}$$^{~~}$ & .38$^{~~}$$^{~~}$\\

& L2R-D SVM$^r$ & .22$^{~~}$[.000] & .48$^{~~}$[.001] & .46$^{~~}$[.004] && .20$^{~~}$[.002] & .34$^{~~}$[.002] & .32$^{~~}$[.005]\\

& GS & .24$^{~~}$[.003] & .51$^{~~}$[.019] & .47$^{~~}$[.003] && .24$^{~~}$[.001] & .40$^{~~}$[.001] & .38$^{~~}$[.020]\\

& Best trained & .25 [.010]$^{~~}$ & .52 [.008]$^{~~}$ & .47 [.009]$^{~~}$ && .22 [.006]$^{~~}$ & .40 [.002]$^{~~}$ & .40 [.012]$^{~~}$\\

& CombSUM & .25 [.002] & .53 [.003] & .50 [.002] && .25 [.005] & .42 [.003] & .40 [.004]\\
\hline

\multirow{3}{*}{\rotatebox[origin=c]{270}{\textbf{SQP}}}
& Trained SQE & .25 [.002]$^{~~}$ & .53 [.007]$^{~~}$ & .49 [.006]$^{~~}$ && .24 [.002]$^{~~}$ & .40 [.016]$^{~~}$ & .39 [.018]$^{~~}$\\

& Random20-RF & .22$^{~~}$$^{~~}$ [.003] & .49$^{~~}$$^{~~}$ [.007] & .46$^{~~}$$^{~~}$ [.004] && .26$^{\vartriangle}$$^{~~}$ [.009] & .41$^{\vartriangle}$$^{~~}$ [.003] & .39$^{\vartriangle}$$^{~~}$ [.008]\\[2pt]

& ERisk-RF & .27$^{\vartriangle}$$^{\uparrow}$ [.005] & .58$^{\vartriangle}$$^{\uparrow}$ [.006] & .55$^{\vartriangle}$$^{\uparrow}$ [.014] && .28$^{\vartriangle}$$^{\uparrow}$ [.017] & .47$^{\vartriangle}$$^{\uparrow}$ [.014] & .43$^{\vartriangle}$$^{\uparrow}$ [.015]\\

& \textbf{ERisk-Cosine} & \textbf{.28}$^{\vartriangle}$$^{\uparrow}$ [.002] & \textbf{.60}$^{\vartriangle}$$^{\uparrow}$ [.002] & \textbf{.57}$^{\vartriangle}$$^{\uparrow}$ [.010] && \textbf{.30}$^{\vartriangle}$$^{\uparrow}$ [.011] & \textbf{.51}$^{\vartriangle}$$^{\uparrow}$ [.002] & \textbf{.49}$^{\vartriangle}$$^{\uparrow}$ [.008]\\
\hline
\multirow{4}{*}{\rotatebox[origin=c]{270}{\textbf{Oracles}}}

& Best conf. & .26 & .54 & .51 && .25 & .42 & .41\\
& Oracle20Erisk & .29$^{~~}$ & .63$^{~~}$ & .61$^{~~}$ && .33$^{~~}$ & .53$^{~~}$ & .52$^{~~}$\\
& Oracle20Random & .24$^{~~}$ & .58$^{~~}$ & .55$^{~~}$ && .31$^{~~}$ & .52$^{~~}$ & .49$^{~~}$ \\
& Oracle & .39$^{~~}$ & .83$^{~~}$ & .80$^{~~}$ && .45$^{~~}$ & .72$^{~~}$ & .69$^{~~}$\\

\end{tabular}\label{tab:trec78WT10Geff_APR_20_20}}
\end{table}
In Table~\ref{tab:trec78WT10Geff_APR_20_20}, we can observe that for these two other ad hoc collections, $E_{Risk}$ models also outperform (statistically significantly) all the baselines. 
We also can draw 
the same conclusion as for Table~\ref{tab:GOV2eff_APR_20_20} regarding baselines where GS is close to the best possible configuration (Best conf.). L2R-D is slightly lower than the best trained and sometimes surpasses the initial BM25. CombSUM and Trained SQE are close one to the other, also close to best trained.

Again, our methods are at the midway point between the best possible configuration of our pool (e.g., on TREC78, nDCG@10 is 0.54 for the best configuration in the $20,000$ ones while our models achieve 0.58 or 0.60, the oracle20Random when using 20 configurations randomly gets 0.58 while the oracle20ERisk is 0.63 when using the 20 configurations selected by $E_{Risk}$.)
With regard to the percentage  effectiveness increase compared to best trained (not reported in the Table), for example, this is  $+15\%$ (resp. $+27\%$) for nDCG@10  on TREC78 (resp. WT10G). 

Separately from GOV2, on these two additional collections, when the 20 configurations are selected randomly rather than selected using $E_{Risk}$, the effectiveness is low both for the L2R model and for the Oracle20Random which would perfectly decide which of the 20 configurations should treat a given query (e.g., nDCG@10 is 0.49 for L2R with Random20 and 0.58 for Oracle20Random, while BM25 is 0.47 on TREC78). These collections are harder than GOV2 and some configurations fail on some queries. Our risk-based methods are very effective on such hard collections.

\subsection{$E_{Risk}$ function for diversity-oriented search }
\label{subsec:EriskDiversity}
The next question we target is whether the model generalizes well on a different task.
Table~\ref{tab:diversity09b12b_APR_20_20}  reports results when considering diversity-oriented search for both the Clueweb09B and Clueweb12B collections, still using the $E_{Risk}$ function with 20 configurations. 

\begin{table*}[ht!]
\caption{Our Risk-based models generalize well on  diversity-oriented search. Here, $k=20$ configurations are used on the Clueweb09B and Clueweb12B collections. $\vartriangle$ (resp. ${\uparrow}$) indicates statistically significant improvement over the L2R documents (resp. grid search (GS)) according to a paired t-test ($p < 0.05$). }{
\centering
\begin{tabular}{@{}l@{\hspace*{0.1cm}}l@{\hspace*{0.1cm}}|@{\hspace*{0.05cm}}l@{\hspace*{0.1cm}} l@{\hspace*{0.1cm}} l@{\hspace*{0.1cm}} c@{}| @{\hspace*{0.05cm}} l@{\hspace*{0.1cm}} l@{\hspace*{0.1cm}} l@{}}
& $E_{Risk}$ & \multicolumn{3}{c}{Clueweb09B} && \multicolumn{3}{c}{Clueweb12B}\\
\hline
& Methods & ERR-IA & $\alpha$-nDCG & NRBP && ERR-IA & $\alpha$-nDCG & NRBP\\
\hline
\multirow{4}{*}{\rotatebox[origin=c]{270}{\textbf{Baselines}}}
& BM25 & .21$^{~~}$ & .18$^{~~}$ & .21 && .35 & .33 & .34\\

& L2R-D. (SVM$^{r}$) & .23 [.001]$^{~~}$ & .33 [.001]$^{~~}$ & .20 [.001]$^{~~}$ && .37 [.013]$^{~~}$ & .46 [.010]$^{~~}$ & .32 [.014]$^{~~}$\\
& GS & .29 [.001]$^{~~}$ & .24 [.002]$^{~~}$ & .28 [.001]$^{~~}$ && .41 [.004]$^{~~}$ & .40 [.002]$^{~~}$ & .42 [.003]$^{~~}$\\
& Best trained & .27 [.022]$^{~~}$ & .22 [.018]$^{~~}$ & .26 [.022]$^{~~}$ && .40 [.019]$^{~~}$ & .37 [.025]$^{~~}$ & .38 [.019]$^{~~}$\\

& {CombSUM} & .27 [.004]$^{~~}$ & .35 [.007]$^{~~}$ & .25 [.004]$^{~~}$ && .46 [.007] & .52 [.008] & .43 [.014]\\
\hline
\multirow{3}{*}{\rotatebox[origin=c]{270}{\textbf{SQP}}}
& Train-SQE & .22 [.009]$^{~~}$ & .19 [.009]$^{~~}$ & .22 [.009]$^{~~}$ && .35 [.009]$^{~~}$ & .33 [.013]$^{~~}$ & .34 [.012]$^{~~}$\\
& Random20-RF & .24$^{~~}$$^{\downarrow}$ [.013] & .20$^{\triangledown}$$^{\downarrow}$ [.008] & .24$^{\vartriangle}$$^{\downarrow}$ [.014] && .38$^{~~}$$^{~~}$ [.012] & .36$^{\triangledown}$$^{~~}$ [.003] & .36$^{~~}$$^{~~}$ [.004]\\
& ERisk-RF & .33$^{\vartriangle}$$^{\uparrow}$ [.011]& .27$^{\triangledown}$$^{\uparrow}$ [.009]& \textbf{.33}$^{\vartriangle}$$^{\uparrow}$ [.010] && .52$^{\vartriangle}$$^{\uparrow}$ [.019]& .51$^{~~}$$^{\uparrow}$ [.015]& .53$^{\vartriangle}$$^{\uparrow}$ [.011]\\

& \textbf{ERisk-Cosine} & .33$^{\vartriangle}$$^{\uparrow}$ [.004] & \textbf{.43}$^{\vartriangle}$$^{\uparrow}$ [.016] & .32$^{\vartriangle}$$^{~~}$ [.007] && \textbf{.58}$^{\vartriangle}$$^{\uparrow}$ [.009] & \textbf{.63}$^{\vartriangle}$$^{\uparrow}$ [.003] & \textbf{.58}$^{\vartriangle}$$^{\uparrow}$ [.005]
\\
\hline
\multirow{4}{*}{\rotatebox[origin=c]{270}{\textbf{Oracles}}}
& Best conf. & .29 & .24 & .29 && .45 & .42 & .44\\
& Oracle20Erisk & .59$^{~~}$ & .48$^{~~}$ & .60$^{~~}$ && .74$^{~~}$ & .72$^{~~}$ & .75$^{~~}$\\
& Oracle20Random & .44$^{~~}$ & .36$^{~~}$ & .44$^{~~}$ && .62$^{~~}$ & .60$^{~~}$ & .63$^{~~}$\\
& Oracle & .67$^{~~}$ & .55$^{~~}$ & .68$^{~~}$ && .80$^{~~}$ & .77$^{~~}$ & .81$^{~~}$\\
\end{tabular}\label{tab:diversity09b12b_APR_20_20}}
\end{table*}

For diversity search, $E_{Risk}$ methods surpass (statistically significantly) L2R documents on both collection for ERR-IA, $\alpha$-nDCG, and NRBP (see Table~\ref{tab:diversity09b12b_APR_20_20}) while $E_{Risk}$ methods surpass (statistically significantly) GS on any measures and for both collections except for NRBP on Clueweb09B collection. They are also higher than the best configuration that GS  finds in the case of Clueweb09B and is close to the best in the case of Clueweb12B. There is no statistically significant differences among the variants of the $E_{Risk}$ models. Erisk-Cosine is better than Erisk-RF for {$\alpha$-nDCG} on Clueweb09B and across all metrics on Clueweb12B.
Like for ad hoc search, Oracle20ERisk and Oracle20Random are obviously lower than Oracle, but Oracle20ERisk is much higher than Oracle20Random. This shows that the selection method is as good as for the ad hoc task. Regarding the third phase of  training, when comparing ERisk-RF and ERisk-Cosine we can see that like for ad hoc, the results are better using the Cosine  query-configuration association (one exception is for NRBP which decreases slightly from RF to Cosine  variants).  
With regards to the percentage of effectiveness increase, for example, ERR-IA increases by $+14\%$ (resp. +41\%) on Clueweb09B (resp. Clueweb12B) compared to GS. We did not report the percentages here to limit the size of the tables.

\subsection{Influence of the number of candidate configurations}
\label{subsec:NbConfig}

The model is designed to select and use a pre-defined number of configurations $k$. The higher the number of configurations, the more complex the meta-system is. It is also likely that the number of configurations will impact on the final results. One could hypothesize that the higher the number of configurations, the better the meta-system effectiveness. However, there is a risk of a non-optimized model if there are too few configurations, and on the other hand there is the risk of overfitting if there are too many configurations compared to the number of examples used in training.

In this section, we  focus on evaluating the ability of the risk functions that we developed to select the appropriate minimal set of configurations. Our objective is not to decide the exact optimized value because it is likely to be dependent on the collection and machine learning method but rather the objective is to get an insight into the evolution of  results according to the number of configurations used.

\begin{figure}[ht!]
\includegraphics[width=0.49\linewidth]{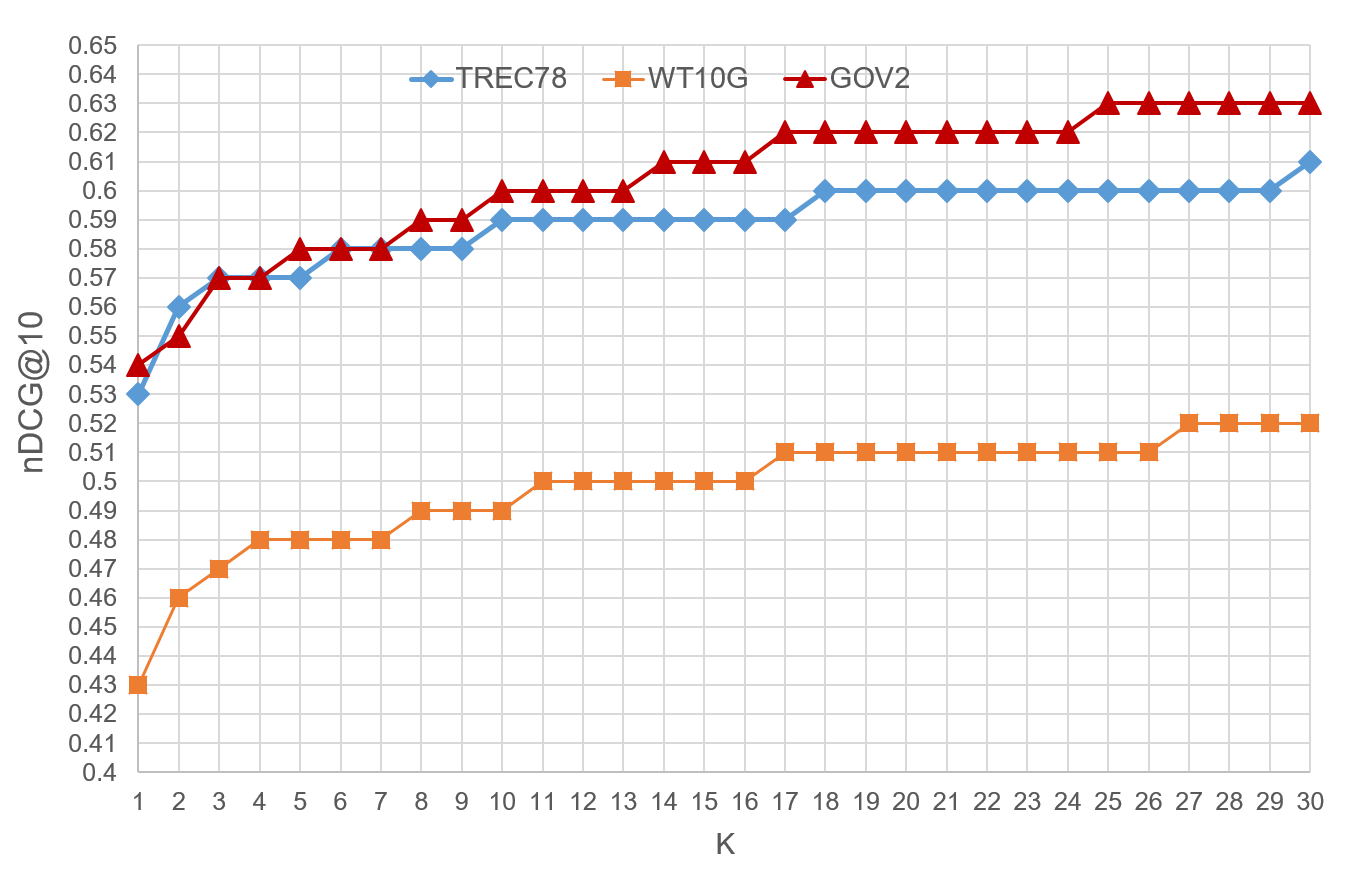}
\includegraphics[width=0.49\linewidth]{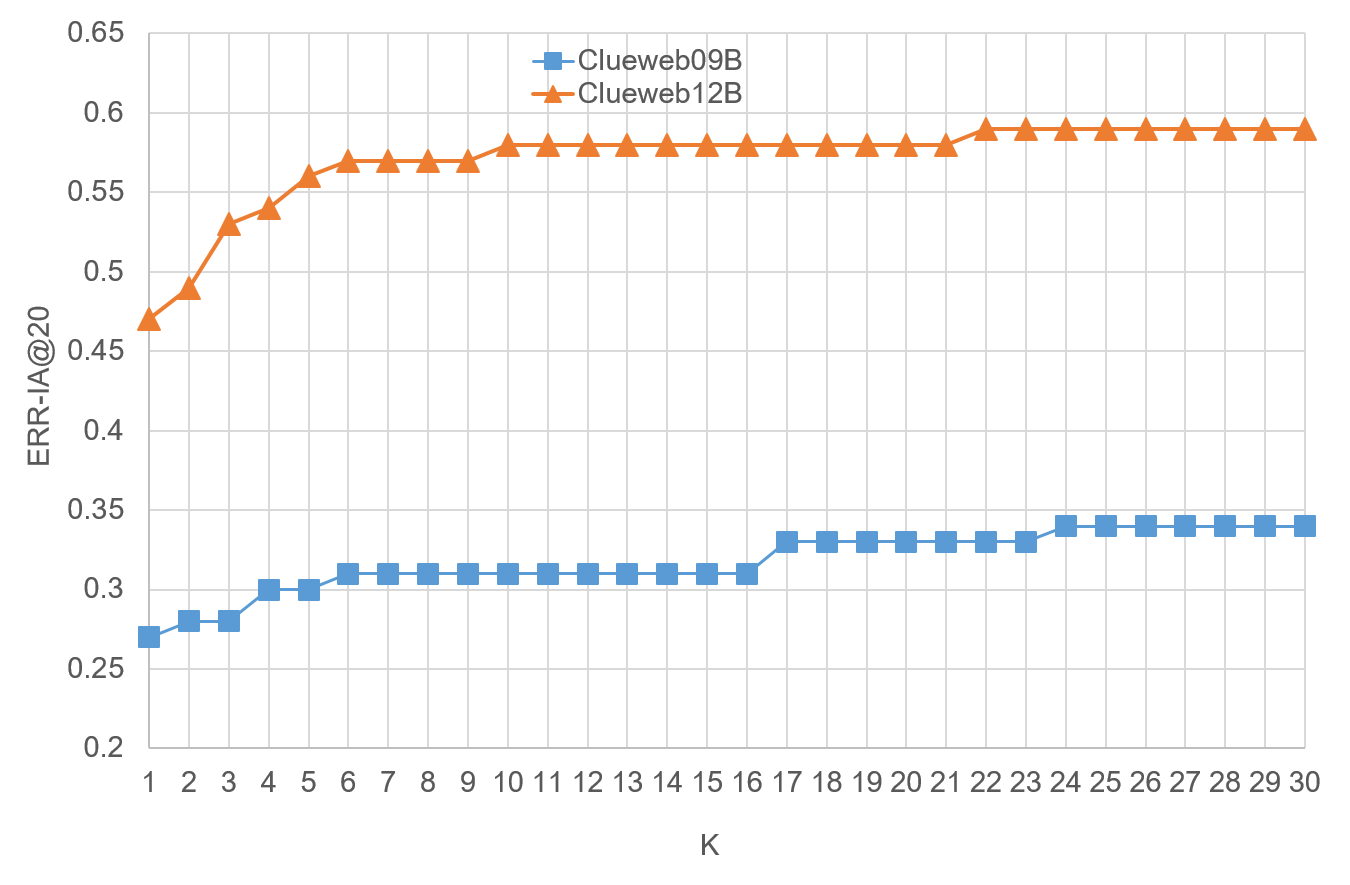}\\
\small{(a) Ad hoc search \hspace{100pt} (b) Diversity-oriented search}
\caption{{Effectiveness grows with $k$, the number of configurations.} The number of configurations $k$ in 1 to 30 (X-axis) and nDCG@10 (Y-axis) for ad hoc collection on (a) or ERR-IA@20 (Y-axis) for diversity collections on (b). {Results are for $E_{Risk}$ function.}}
\label{fig:kvariants}
\end{figure}

For this, we report in Figure~\ref{fig:kvariants} the results with the $E_{Risk}$ function on the three collections for ad hoc and on  two collections for diversity, for different values of $k$. $k$ is the number of configurations that are selected by $E_{Risk}$ from among the pool of configurations  that are then  candidates for  selection in the third phase. 
We use a single query trial since we have shown in Section~\ref{subsec:EriskAdhoc} that the results are stable and robust.

We can observe from Figure~\ref{fig:kvariants} that effectiveness increases with the number of configurations used, growing in increments. We can observe some stages on the curves which mean that an added configuration does not increase overall performance. Considering the $E_{Risk}$ iterative process, that means that either the newly added configuration just improves slightly the effectiveness on one or very few queries which is not visible once averaged over the query set, or that this configuration is not selected in the third phase of the process. In that latter case, that means that the configuration would be useful if the query-configuration matching was appropriate.

\subsection{$E_{Risk}$ vs $N_{Risk}$ functions}
\label{subsec:2func}

In Section~\ref{sec:risk}, we developed two risk functions that do not have the same objectives. While $E_{Risk}$ aims at optimizing the effectiveness values, $N_{Risk}$ aims at optimizing the number of queries for which effectiveness will be improved. As a result, overall effectiveness for $N_{Risk}$ cannot be higher than that for $E_{Risk}$, but the expected advantage of $N_{Risk}$ over $E_{Risk}$ is that it minimizes the number of queries that have poor results and is thus a user-oriented function. Indeed overall effectiveness can be improved because  the results of ``easy" queries have been improved while search engines should also care about the problem of ``hard" queries. In the past, Mizzaro and Roberts mentioned that ``a system that wants to be effective in TREC needs to be effective on easy topics"~\cite{mizzaro2007hits}. $N_{Risk}$ tackles this problem by trying to improve all the queries, considering it is more appropriate to improve slightly a lot of query results rather than improving a lot a few queries that may be already well answered by the system.

\begin{table*}[ht!]
\caption{$N_{Risk}$ average effectiveness is slightly over that of $E_{Risk}$. GOV2 (ad hoc) and Clueweb12B (diversity) collections are used. Effectiveness and standard deviations are first reported, then the number of queries for which the results are increased ($\blacktriangle$) or decreased ($\triangledown$) compared to GS. $\vartriangle$ (resp. ${\uparrow}$) indicates statistically significant effectiveness improvement over the L2R document (resp. grid search (GS)) according to a paired t-test ($p < 0.05$).}{
\centering
\begin{tabular}{@{}l l@{\hspace*{0.1cm}} | r@{\hspace*{0.2cm}} | r@{\hspace*{0.2cm}} | r@{\hspace*{0.2cm}} c@{}| l@{}}
& $N_{Risk}$& \multicolumn{3}{c}{GOV2} \\
\hline
& & AP & nDCG@10 & P@10 \\
\hline
\multirow{4}{*}{\rotatebox[origin=c]{270}{\textbf{Baselines}}}
& BM25 & .27 $\blacktriangle$031 - $\triangledown$118 &
.46 $\blacktriangle$048 - $\triangledown$85 &
.54 $\blacktriangle$31 - $\triangledown$78 \\

& L2R-D. (SVM$^{r}$) & .28 $\blacktriangle$035 - $\triangledown$113 & .49 $\blacktriangle$052 - $\triangledown$78 & .57 $\blacktriangle$37 - $\triangledown$65 \\

& Best trained. & .35 $\blacktriangle$065 - $\triangledown$039 &
.49 $\blacktriangle$059 - $\triangledown$72 & .59 $\blacktriangle$47 - $\triangledown$58\\

& CombSUM & .36 $\blacktriangle$096 - $\triangledown$051 & .54 $\blacktriangle$072 - $\triangledown$56 & .65 $\blacktriangle$44 - $\triangledown$36 \\ 

& GS &.35$^{~~} \hspace{50pt}$ & .52$^{~~}\hspace{45pt}$ & .62$^{~~}\hspace{40pt}$\\
\hline

\multirow{3}{*}{\rotatebox[origin=c]{270}{\textbf{SQP}}}
& Random20 & .33$^{\vartriangle}$$^{\downarrow}$ $\blacktriangle$047 - $\triangledown$101 & .48$^{~~}$$^{\downarrow}$ $\blacktriangle$053 - $\triangledown$80 & .58$^{~~}$$^{~~}$ $\blacktriangle$044 - $\triangledown$70\\
& NRisk-RF & .38$^{\vartriangle}$$^{\uparrow}$ $\blacktriangle$103 - $\triangledown$045 & .60$^{\vartriangle}$$^{\uparrow}$ $\blacktriangle$099 - $\triangledown$40 & .73$^{\vartriangle}$$^{\uparrow}$ $\blacktriangle$083 - $\triangledown$17 \\

& \textbf{NRisk-Cosine} & \textbf{.45}$^{\vartriangle}$$^{\uparrow}$ $\blacktriangle$147 - $\triangledown$001 & \textbf{.75}$^{\vartriangle}$$^{\uparrow}$ $\blacktriangle$139 - $\triangledown$00 & \textbf{.84}$^{\vartriangle}$$^{\uparrow}$ $\blacktriangle$107 - $\triangledown$00\\
\hline
& Best conf. & .35\hspace{5pt} $\blacktriangle$079 - $\triangledown$054 & .49\hspace{5pt} $\blacktriangle$054 -  $\triangledown$54 & .59\hspace{5pt} $\blacktriangle$16 - $\triangledown$16 \\
\\[5pt]

&$N_{Risk}$  & \multicolumn{3}{c}{Clueweb12B} \\
\hline
&&ERR-IA & $\alpha$-nDCG & NRBP\\
\hline
\multirow{4}{*}{\rotatebox[origin=c]{270}{\textbf{Baselines}}}
& BM25 & .35 $\blacktriangle$38 - $\triangledown$52 &
.33 $\blacktriangle$47 - $\triangledown$41 &
.34 $\blacktriangle$27 - $\triangledown$61 \\

& L2R-D. (SVM$^{r}$) & .37 $\blacktriangle$45 - $\triangledown$48 & .46 $\blacktriangle$59 - $\triangledown$33 & .32 $\blacktriangle$36 - $\triangledown$55\\

& Best trained. &  .40 $\blacktriangle$32 - $\triangledown$34 & .37 $\blacktriangle$50 - $\triangledown$42 & .37 $\blacktriangle$50 - $\triangledown$42\\
& CombSUM & .46 $\blacktriangle$58 - $\triangledown$33 & .52 $\blacktriangle$66 - $\triangledown$23 & .43 $\blacktriangle$52 - $\triangledown$38\\
& GS & .41$^{~~}\hspace{42pt}$ & .40$^{~~}\hspace{42pt}$ & .42$^{~~}\hspace{42pt}$\\

\hline
\multirow{3}{*}{\rotatebox[origin=c]{270}{\textbf{SQP}}}
& Random20 & .38$^{~~}$$^{~~}$$\blacktriangle$35 - $\triangledown$49 & .37$^{\triangledown}$$^{~~}$$\blacktriangle$45 - $\triangledown$43 & .38$^{~~}$$^{~~}$$\blacktriangle$44 - $\triangledown$47\\[2pt]
& NRisk-RF & .38$^{~~}$$^{~~}$$\blacktriangle$31 - $\triangledown$56 & .36$^{\triangledown}$$^{~~}$$\blacktriangle$46 - $\triangledown$50 & .37$^{~~}$$^{~~}$$\blacktriangle$49 - $\triangledown$43\\ 

& \textbf{NRisk-Cosine} & \textbf{.67}$^{\vartriangle}$$^{\uparrow}$ $\blacktriangle$94 - $\triangledown$5 & \textbf{.70}$^{\vartriangle}$$^{\uparrow}$ $\blacktriangle$92 - $\triangledown$7 & \textbf{.67}$^{\vartriangle}$$^{\uparrow}$ $\blacktriangle$89 - $\triangledown$9\\
\hline
& Best conf. & .45 \hspace{5pt}$\blacktriangle$59 - $\triangledown$35 & .42\hspace{5pt} $\blacktriangle$68 - $\triangledown$26 & .44 \hspace{5pt}$\blacktriangle$55 - $\triangledown$39\\
\end{tabular}\label{tab:Nrisk}}

\end{table*}

Table~\ref{tab:Nrisk} describes the results for $N_{Risk}$. We report the average effectiveness and standard deviations as well as the number of queries for which the models improve/decrease the effectiveness of the query compared to grid search  (GS, the best baseline, see Tables~\ref{tab:GOV2eff_APR_20_20} and \ref{tab:trec78WT10Geff_APR_20_20}). 

Compared to $E_{Risk}$, we expect $N_{Risk}$ to be slightly less efficient on average but to improve more queries than baselines. The difference between $E_{Risk}$ and $N_{Risk}$ in terms of effectiveness is small (e.g., -3\% on GOV2 for nDCG@10 and P@10) and it is slightly larger for diversity-oriented search.

$N_{Risk}$ targets the number of improved queries and improves more queries than $E_{Risk}$ does. For example, considering Risk-RF on GOV2 with P@10, when compared with GS, there are 107 queries that are improved  with $N_{Risk}$, while there are 80 with $E_{Risk}$. More surprisingly, the overall effectiveness is also higher with $N_{Risk}$ than with $E_{Risk}$ which however targets effectiveness: on GOV2, P@10 is 0.76 for $E_{Risk}$ and 0.84 for $N_{Risk}$, nDCG@10 is 0.62 for $E_{Risk}$, and 0.75 for $N_{Risk}$. The same holds for the Clueweb12B collection. Most of the comments we made on $E_{Risk}$ regarding  comparisons between the baselines among them and with on the one hand selective methods and on the other hand Oracle, also hold for $N_{Risk}$. The BM25 baseline is the lowest. Learning to rank improves that baseline but a single configuration that involved training is better. Selective query processing are the best and among them the Cosine  version of our risk-based method is the best performing and surpasses the best single configuration an Oracle would choose. The Cosine  version is also the best in terms of improved queries; more than 3/4 of the queries are improved.

Since these two objective functions $E_{Risk}$ and $N_{Risk}$ do not target the same goal, they could also be combined. We keep this for future work. 

\subsection{Impact of Shallow vs. Deep Relevance Judgments}
\label{Sub:MS-MARCO}
In the previous series of experiments, we considered collections for which the number of judged documents (positive and negative) is rather high (see Table~\ref{tab:collection}). For example, on the GOV2 collection, on average there are  902 assessed documents per query, 179 of which are marked as relevant. In this section, we consider the MS MARCO document ranking collection\footnote{\url{https://microsoft.github.io/MS-MARCO-Document-Ranking-Submissions/leaderboard/}}, which has shallow relevance judgments. We consider the development set for which there are 5,193 queries with only one relevant document per query and no negative judgments. This collection thus offers  a completely different evaluation context for ad hoc retrieval. 

We first report the results when using the usual ad hoc metrics (see Table~\ref{tab:MS-MARCO-Dev_20_20})\footnote{MAP is replaced here by mean reciprocal rank (MRR) since there is a single known relevant document. Reciprocal rank is defined as 1/rank where rank is the rank of the relevant document. MRR is RR averaged over the topics}. The results for $E_{Risk}$ are consistent with those found for  other ad hoc collections and  diversity collections. Specifically, $E_{Risk}$ combined with a Cosine  similarity in the third phase is very effective. It is slightly better than our other version of the risk function. The differences compared to  other trained methods is less impressive  for the other collections, but MS-MARCO evaluation has biases. However, the better performance of the $E_{Risk}$ over both learning to rank documents and the best trained single configuration is statistically significant. 

\begin{table}[!ht]
\caption{Our selective query processing model is the most effective when shallow relevance judgments are available in terms of RR, 
nDCG@10 and P@10 for MS MARCO collection. Results are averaged based on 3 draws and two test folds. Absolute values and the standard deviations in square brackets are reported. For our methods (ERisk-xx) with $k=$20 candidate configurations, $\vartriangle$ (resp. ${\uparrow}$) indicates statistically significant improvements over the L2R document (resp. Best trained) according to a paired t-test ($p < 0.05$) with Bonferroni correction.}{
\centering
\begin{tabular}{@{}l@{\hspace*{0.2cm}} l@{\hspace*{0.1cm}}|@{\hspace*{0.1cm}}l@{\hspace*{0.1cm}} l@{\hspace*{0.1cm}} c@{}}
\multicolumn{2}{l}{$E_{Risk}$} & \multicolumn{3}{c}{MS MARCO }\\
\hline
& & \multicolumn{3}{c}{Absolute values}\\
\cline{3-5}
& Methods & RR & nDCG@10 & P@10\\
\hline
\multirow{4}{*}{\rotatebox[origin=c]{270}{\textbf{Baselines}}}
& BM25 & .27 & .32 & .05\\
& L2R-D SVM$^r$ & .28 [.000] & .33 [.000] & .05 [.000]\\
& Best trained & .27 [.002] & .32 [.002] & .05 [.000]\\
& CombSUM & .28 [.001] & .33 [.001] & .05 [.000]\\

\hline
\multirow{3}{*}{\rotatebox[origin=c]{270}{\textbf{SQP}}}
& Trained SQE & .28$^{~~}$$^{~~}$ [.000]$^{~~}$ & .32$^{~~}$$^{~~}$ [.001]$^{~~}$ & .05$^{~~}$$^{~~}$ [.000]$^{~~}$\\
& Random20-SVM$^{r}$ & .22$^{\triangledown}$$^{\downarrow}$ [.045] & .27$^{\triangledown}$$^{\downarrow}$ [.055] & .05$^{\triangledown}$$^{\downarrow}$ [.004]\\
& ERisk-SVM$^{r}$ & .28$^{~~}$$^{~~}$ [.000] & .32$^{~~}$$^{~~}$ [.001] & .05$^{~~}$$^{~~}$ [.000]\\
& \textbf{ERisk-Cosine} & \textbf{.29}$^{\vartriangle}$$^{\uparrow}$ [.006] & \textbf{.34}$^{\vartriangle}$$^{\uparrow}$ [.005] & \textbf{.05}$^{\vartriangle}$$^{\uparrow}$ [.000]\\
\hline
\multirow{4}{*}{\rotatebox[origin=c]{270}{\textbf{Oracles}}}
& Best conf. & .28 & .32 & .05\\
& Oracle20Random & .44 & .49 & .07\\
& Oracle20ERisk & .33 & .38 & .06\\
& Oracle & .70 & .73 & .08\\
\end{tabular}\label{tab:MS-MARCO-Dev_20_20}}
\end{table}

\subsection{Insights into the results}
\label{sub:deep}

\vspace{6pt}
\noindent 
\textbf{Insights on the selected configurations}

\begin{figure}[htbp]
     \centering
     \begin{subfigure}[b]{0.49\textwidth}
        \centering
        \includegraphics[width=6.8cm,height=8.7cm]{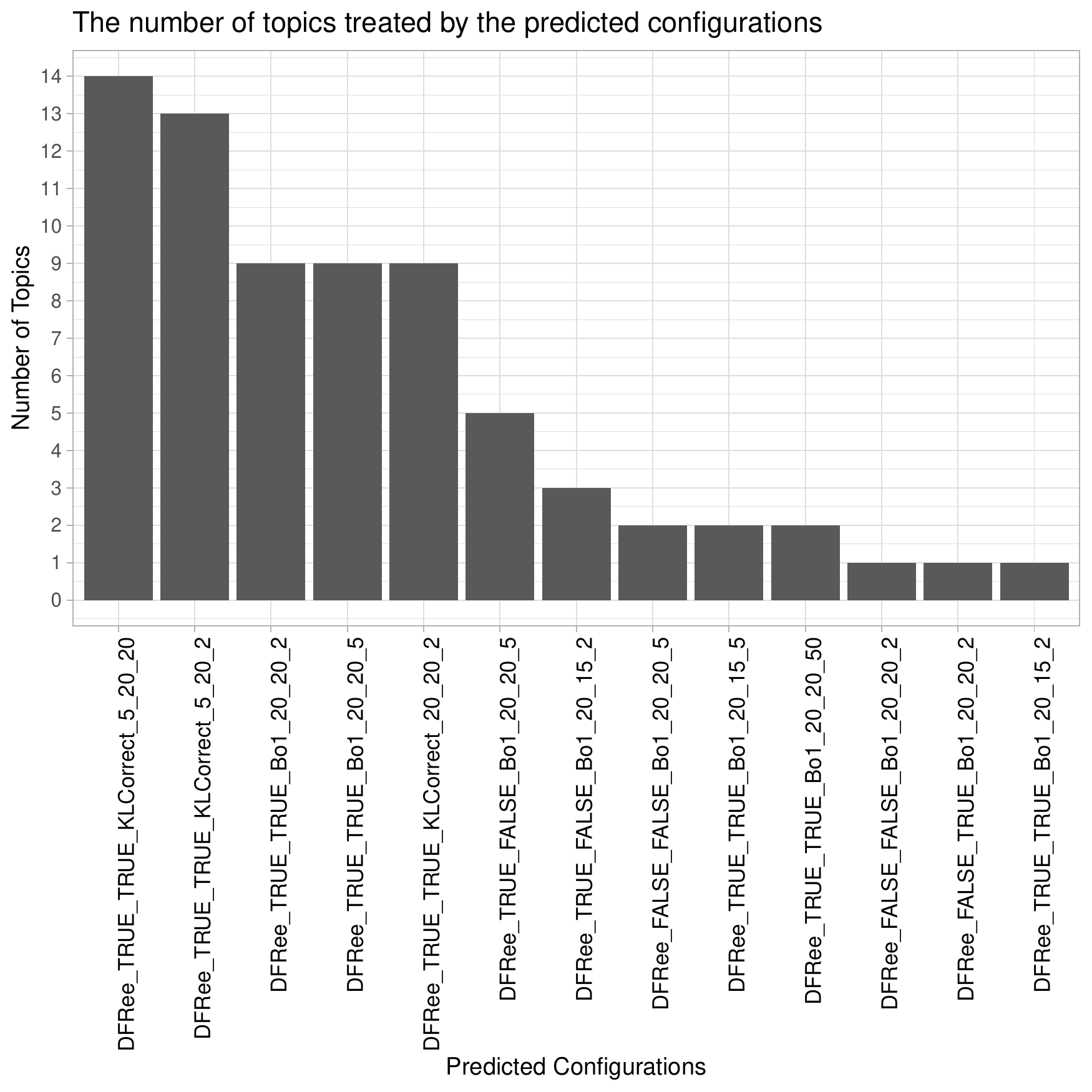}
        \caption{ERisk-Cosine (AP)}
        \label{fig:erisk-erria20-clueweb12bdiv}
     \end{subfigure}
     \begin{subfigure}[b]{0.49\textwidth}
         \centering
         \includegraphics[width=6.8cm,height=8.7cm]{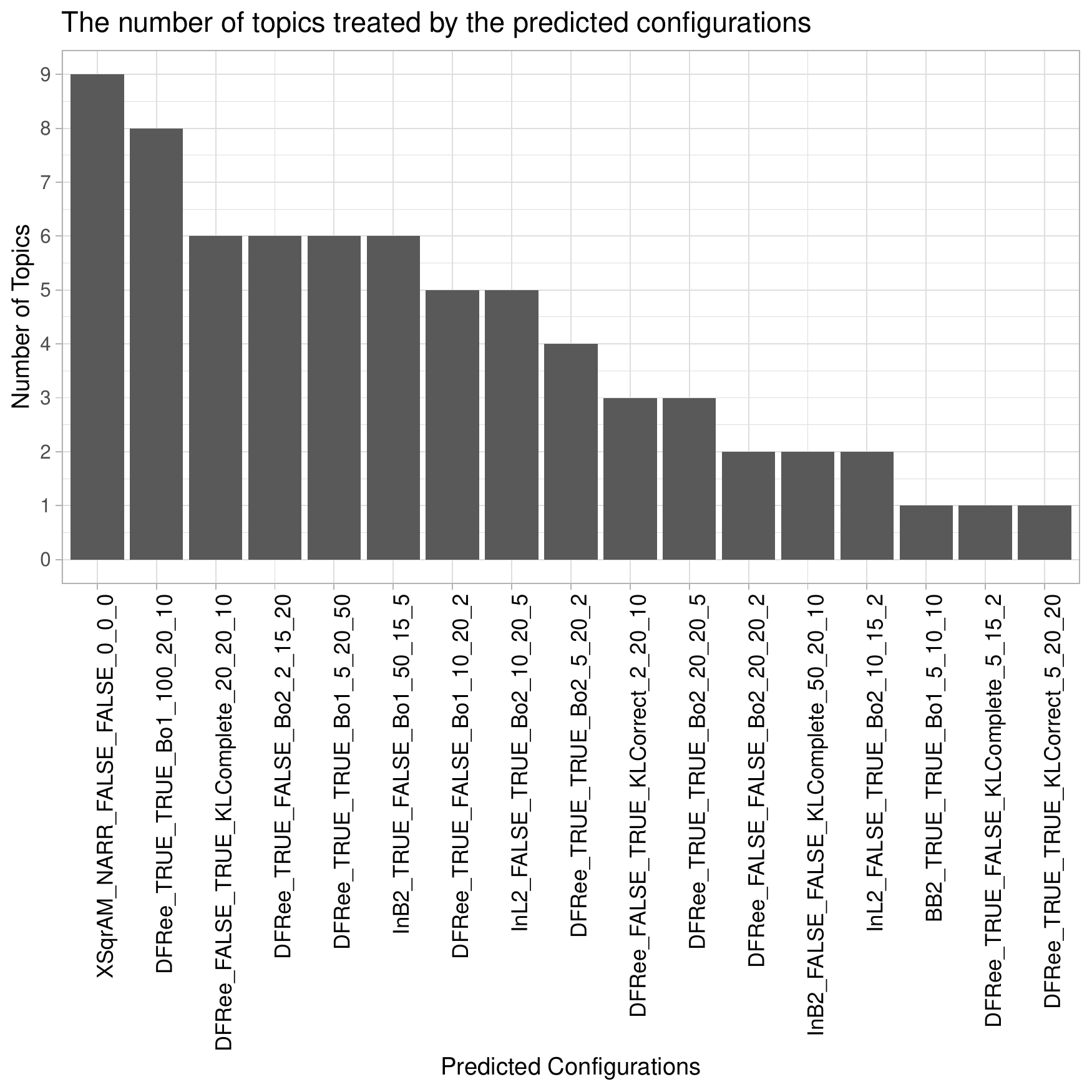}
         \caption{NRisk-Cosine (AP)}
         \label{fig:random-erria-clueweb12bdiv}
     \end{subfigure}
     \caption{Number of queries for which a given configuration is predicted. Bar plots for the 20  configurations chosen by 
     (a) $E_{Risk}$-Cosine, and (b) $N_{Risk}$-Cosine; GOV2 using AP.} 
    \label{fig:BarplotGov2-ERisk-Cosine-vs-RF}
\end{figure}

While our risk-sensitive functions  choose some configurations as candidates for the meta-system, we are not sure if they are  selected for the test/unseen queries. Figure~\ref{fig:BarplotGov2-ERisk-Cosine-vs-RF} shows the distribution of the test queries according to the candidate configurations that the meta-system chose on the test queries after training on the training queries. We  randomly select one of the trials of our three-cross validation for 20 configurations. We plot AP for the GOV2 collection but the others have the same general shape.

We can see that some configurations are not selected. There are 20 configurations in total but here 13 and 17 configurations were selected for at least one query. Some configurations are used for just one query, but several configurations are chosen for several queries. For example, the most used configuration is selected for 14 queries with $E_{Risk}$ and for 8 queries with $N_{Risk}$ on GOV2 and AP measure. Deeper analysis could be conducted in order to 
better optimize the number of configurations. This optimization could also take into consideration the query processing time (an automatically expanded query is more expansive to process than an initial query for example).

\begin{figure}[htbp]
     \centering
     \begin{subfigure}[b]{0.49\textwidth}
         \centering
         \includegraphics[width=6.8cm,height=8.7cm]{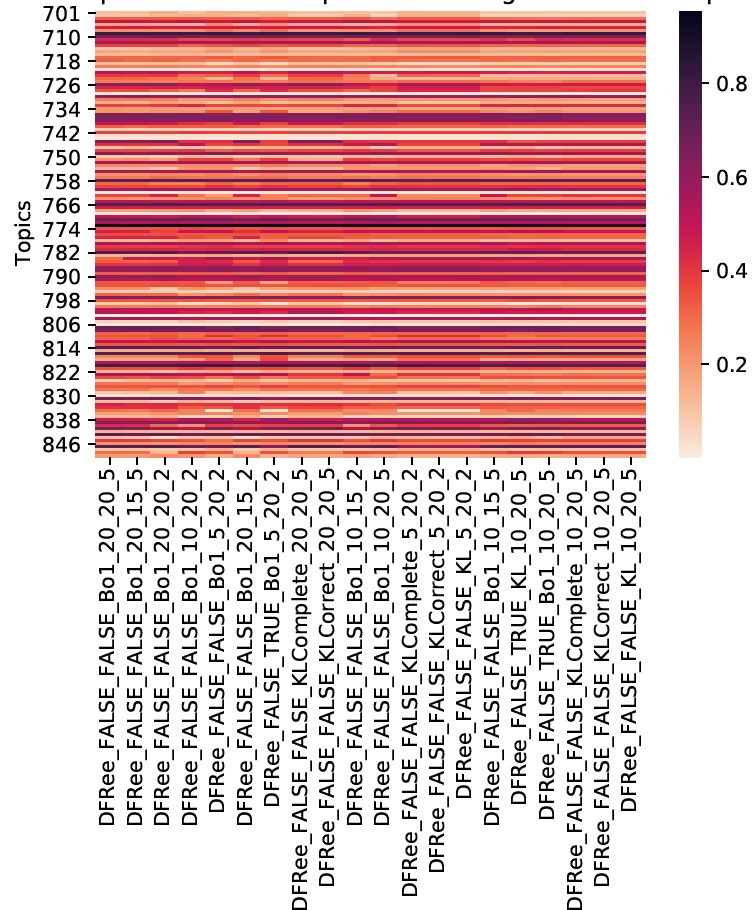}
         \caption{Erisk}
         \label{fig:erisk-map-gov2}
     \end{subfigure}
     \begin{subfigure}[b]{0.49\textwidth}
         \centering
         \includegraphics[width=6.8cm,height=8.7cm]{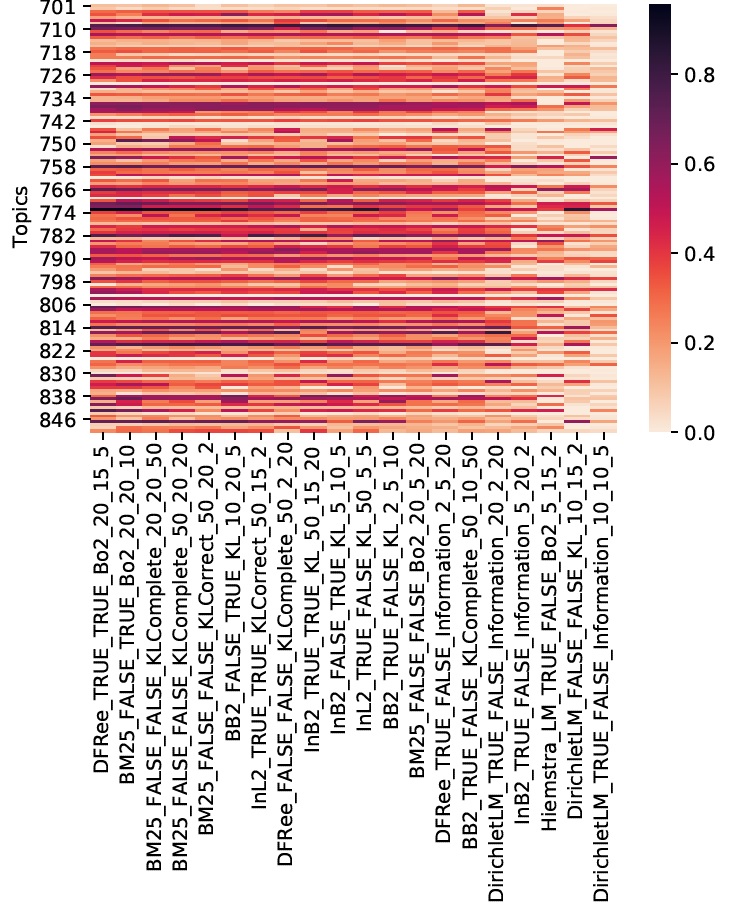}
         \caption{Random}
         \label{fig:random-map-gov2}
     \end{subfigure}
     \caption{{Configurations selected by $E_{Risk}$ are more robust to treat any query.}
     Heatmap plots for the 20 selected configurations selected by (a) $E_{Risk}$, (b) randomly; GOV2 using AP.} 
        \label{fig:HeatGov2Clueweb12b}
\end{figure}

Figure~\ref{fig:HeatGov2Clueweb12b} provides another interesting insight into the selected configurations. We observe that the risk of drastically decreasing effectiveness is much higher in random selection than $E_{Risk}$ if the association of configurations with the query is wrong (see Figure~\ref{fig:HeatGov2Clueweb12b}).
   
In this figure the heatmap on the left side  shows the effectiveness (AP) of 20 configurations in the case of $E_{Risk}$ while the heatmap on the right side  is the one with random selection. We can see that for GOV2, the $E_{Risk}$ is much more effective in general on the queries (the overall heatmap is darker 
in Figure~\ref{fig:HeatGov2Clueweb12b}). The same type of results were obtained for the other collections. Even if the system does not associate the best configuration with the query, the effectiveness will not drastically decrease in the case of $E_{Risk}$ while the best match is more crucial in the case of random selection. Indeed, while the perfect match using Oracle works almost the same for both random and  $E_{Risk}$ selections, it is not the case for the automatic matching.

\vspace{6pt}
\noindent
\textbf{{Impact of relevance judgments in the Top-n}}

The relevance judgments in each collection (qrels) include relevant or irrelevant documents except for MS MARCO (see Figure~\ref{tab:collection}). Thus, given a query, the \textit{top-n} retrieved documents from a corpus may include unjudged documents, which could be either relevant or irrelevant for the underlying query. We analyzed the coverage of judged (qrels) documents in the \textit{top-n} documents retrieved from the corpus by the $20,000$ configurations and the 20 candidate configurations (E$_{Risk}$) corresponding to each collection. 

This coverage gives insights with regard to the reliable use of a performance metric at a particular cutoff. When comparing the coverage in between the initial configuration pool and the subset that the risk-sensitive function chooses, we can see the potential of the configurations. 

We report the percentage of the top-n documents retrieved by the configurations that had been judged in Table~\ref{tab:configuration_qrels}. The percentage of the top-n documents retrieved by the configurations that had been judged is computed as follows: $$((\sum_{j=1}^{C} \sum_{i=1}^{|\mathcal{Q}|}  \frac{|D^{n}_{i,j} \cap Qrel_{i}|}{\textit{top-n}})*100)/(C*|\mathcal{Q}|)$$
where $C$ is the set of configurations, $\mathcal{Q}$ is the entire set of queries, $\frac{|D^{n}_{i,j} \cap Qrel_{i}|}{\textit{top-n}}$ is the ratio of the number of judged documents in the \textit{top-n} documents retrieved for the $i$-th query of the $j$-th configuration to the \textit{top-n} cutoff rank.

From Table~\ref{tab:configuration_qrels}, we can observe that when considering the full set of configurations, they can retrieve more than 80\% of the judged documents in the top-10 rank for the TREC78, WT10G, and GOV2 collections. This is only 3.7\% for the MS MARCO collection but this low coverage is expected  because queries have only one judged (relevant) document. Regarding diversity collections, the configurations can retrieve 54\% of the judged documents in the top-10 rank for Clueweb09B and  46\% for Clueweb12B. This latter collection contains many noisy documents. Additionally, we report the statistics of judged documents retrieved for the 20 candidate configurations selected by the E$_{Risk}$ function. We can see that these 20 candidate configurations generally bring more relevant documents 
in the \textit{top-n}, the only exception being for TREC78. We will see in the next sub-section that this is likely to be due to a better selection of relevant documents and not due to  filtering out non-judged documents.

Although the percentage of relevant documents is relatively high in the top-10 cutoff ranks for the 20 candidate configurations, our proposed approach should  still work at higher depth rank positions, for example, top-50 or top-100 ranks when we have a few relevant documents. We keep the analysis of the impact of higher depth ranks for future work.

\begin{table}[ht!]
\caption{Percentage of retrieved documents that are included  by the configurations and by the candidate configurations (at top-$n$ rank) that have qrels and have been judged.
}
{
\begin{tabular}{@{} l@{\hspace*{0.5cm}} l @{\hspace*{0.5cm}} c @{\hspace*{0.5cm}} c @{\hspace*{0.5cm}} r@{\hspace*{0.5cm}} r@{\hspace*{0.5cm}} r@{\hspace*{0.5cm}} r@{}}
& Collection & \#Conf. & Top-10 & Top-20 & Top-30 & Top-50 & Top-100\\
\hline
\multirow{3}{1.2cm}{Ad hoc}
& TREC78 & 25520 & 89\% & 88\% & 86\% & 84\% & 79\%\\
& TREC78 ($E_{Risk}$) & 20 & 80\% & 80\% & 79\% & 78\% & 76\%\\[3pt]
& WT10G & 25520 & 81\% & 78\% & 76\% & 74\% & 69\%\\
& WT10G ($E_{Risk}$) & 20 & 99\% & 99\% & 99\% & 98\% & 95\%\\[3pt]
& GOV2 & 25520 & 86\% & 83\% & 81\% & 78\% & 72\%\\
& GOV2 ($E_{Risk}$) & 20 & 99\% & 98\% & 97\% & 96\% & 92\%\\[3pt]
& MS MARCO & 5080 & 3.7\% & 2.3\% & 1.7\% & 1.1\% & 0.6\%\\
& MS MARCO ($E_{Risk}$) & 20 & 5.2\% & 3.1\% & 2.3\% & 1.5\% & 0.8\%\\
\hline
\multirow{2}{*}{Diversity}
& Clueweb09B & 25520 & 54\% & 48\% & 43\% & 36\% & 27\%\\
& Clueweb09B ($E_{Risk}$) & 20 & 68\% & 61\% & 55\% & 47\% & 35\%\\[3pt]
& Clueweb12B & 25520 & 46\% & 39\% & 35\% & 29\% & 20\%\\
& Clueweb12B ($E_{Risk}$) & 20 & 67\% & 54\% & 51\% & 41\% & 29\%\\
\end{tabular}\label{tab:configuration_qrels}}
\end{table}

\vspace{6pt}
\noindent
\textbf{Uncertainty of performance}

To observe the approximate uncertainty of performance, we consider the RBP metric~\cite{rbp-moffat-zobel} and report the results for both the MS MARCO (shallow)  and TREC78 (deep) collections. We also report the results of the RBP lower bound as well as the residual error, RBP$_{+Error}$ (see Table~\ref{tab:MSMARCO-dev_rbp_10}). This will give us insights into whether our methods just filter out unjudged documents from the retrieved document list.

From the right side and upper-part of Table~\ref{tab:MSMARCO-dev_rbp_10}, we can observe the residual error RBP$_{+Error}$, which highlights the RBP upper bound if we consider the unjudged documents in the retrieved list as judged and relevant. In the case of the shallow judgments collection (MS MARCO top part of the Table), the residual error is very high, the reason being that the qrels has only one judged and relevant document per query and in other words, 99\% of the documents are unjudged. We can observe that the residual error is very high whatever  method used. Even Oracle with $5,080$ configurations has a high residual error and MS MARCO is also known for having biases~\cite{gupta2022survivorship}. 

\begin{table}[!ht]
\caption{The experimental results on MS MARCO and TREC78 reveal a similar trend like other collections in terms of the ranked biased precision (RBP) metric. Results are averaged using RBP for persistence values at 0.5, 0.8, and 0.95, based on 3 draws and two test folds. Absolute values and  standard deviations in square brackets are reported. On the left-side, RBP values are reported while residual errors are reported on the right-side. For our methods (ERisk-xx) with $k=20$ candidate configurations, $\vartriangle$ (resp. ${\uparrow}$) indicates statistically significant improvement over the L2R document (resp. Best trained) according to a paired t-test ($p < 0.05$) with Bonferroni correction. We use the same notation as used in Table~\ref{tab:GOV2eff_APR_20_20} 
}{
\centering
\centering
\begin{tabular}{@{}l@{\hspace*{0.2cm}} l@{\hspace*{0.1cm}}|@{\hspace*{0.05cm}}l@{\hspace*{0.1cm}} l@{\hspace*{0.1cm}} l@{\hspace*{0.1cm}} c@{}| @{\hspace*{0.05cm}} l@{\hspace*{0.1cm}} l@{\hspace*{0.1cm}} l@{}}
& $E_{Risk}$ & \multicolumn{7}{c}{MS MARCO}\\
\hline
& & \multicolumn{3}{c}{RBP} && \multicolumn{3}{c}{RBP$_{+Error}$}\\
\hline
& Methods & .50 & .80 & .95 && .50 & .80 & .95\\
\hline
\multirow{4}{*}{\rotatebox[origin=c]{270}{\textbf{Baselines}}}
& BM25 & .09 & .06 & .02 && .91 & .94 & .98\\[2pt]
& L2R-D SVM$^r$ & .12$^{~~}$$^{~~}$$^{~~}$ [.000] & .07$^{~~}$$^{~~}$$^{~~}$ [.000] & .03$^{~~}$$^{~~}$$^{~~}$ [.000] && .88 [.000] & .93 [.000] & .97 [.000]\\
& Best trained & .13$^{~~}$$^{~~}$$^{~~}$ [.001] & .07$^{~~}$$^{~~}$$^{~~}$ [.000] & .03$^{~~}$$^{~~}$$^{~~}$ [.000] && .88 [.000] & .93 [.000] & .97 [.000]\\
& CombSUM & .12$^{~~}$$^{~~}$$^{~~}$ [.000] & .07$^{~~}$$^{~~}$$^{~~}$ [.000] & .03$^{~~}$$^{~~}$$^{~~}$ [.000] && .88 [.000] & .93 [.000] & .97 [.000]\\
\hline

\multirow{3}{*}{\rotatebox[origin=c]{270}{\textbf{SQP}}}

& Random-SVM$^{r}$ & .09$^{~~}$$^{~~}$$^{~~}$ [.022] & .06$^{~~}$$^{~~}$$^{~~}$ [.005] & .02$^{~~}$$^{~~}$$^{~~}$ [.003] && .91 [.023]$^{~~}$ & .94 [.006]$^{~~}$ & .98 [.006]$^{~~}$\\

& ERisk-SVM$^{r}$ & .12$^{~~}$$^{~~}$$^{~~}$ [.001] & .07$^{~~}$$^{~~}$$^{~~}$ [.000] & .03$^{~~}$$^{~~}$$^{~~}$ [.000] && .88 [.000]$^{~~}$ & .93 [.000]$^{~~}$ & .97 [.000]$^{~~}$\\

& \textbf{ERisk-Cosine} & .13$^{\vartriangle}$$^{\uparrow}$ [.010] & .07$^{~~}$$^{~~}$$^{~~}$ [.000] & .03$^{~~}$$^{~~}$$^{~~}$ [.000] && .92 [.006] & .95 [.006] & .98 [.000]\\
\hline

\multirow{4}{*}{\rotatebox[origin=c]{270}{\textbf{Oracles}}}
& Best conf. & .12 & .07 & .03 && .88 & .93 & .97\\
& Oracle20Random & .22 & .11 & .04 && .78 & .89 & .96\\
& Oracle20ERisk & .14 & .08 & .03 && .86 & .92 & .97\\
& Oracle & .38 & .17 & .05 && .56 & .75 & .86\\

\\[5pt]

& $E_{Risk}$ & \multicolumn{7}{c}{TREC78}\\
\hline
& & \multicolumn{3}{c}{RBP} && \multicolumn{3}{c}{RBP$_{+Error}$}\\
\hline
& Methods & .50 & .80 & .95 && .50 & .80 & .95\\
\hline
\multirow{4}{*}{\rotatebox[origin=c]{270}{\textbf{Baselines}}}
& BM25 & .49 & .45 & .32 && .01 & .01 & .01\\[2pt]
& L2R-D SVM$^r$ & .51$^{~~}$$^{~~}$$^{~~}$ [.006] & .46$^{~~}$$^{~~}$$^{~~}$ [.000] & .34$^{~~}$$^{~~}$$^{~~}$ [.001] && .00 [.000]$^{~~}$ & .00 [.000]$^{~~}$ & .01 [.006]$^{~~}$\\
& Best trained & .56$^{~~}$$^{~~}$$^{~~}$ [.001] & .49$^{~~}$$^{~~}$$^{~~}$ [.006] & .35$^{~~}$$^{~~}$$^{~~}$ [.004] && .00 [.000]$^{~~}$ & .01 [.006]$^{~~}$ & .02 [.006]$^{~~}$\\
& CombSUM & .59$^{~~}$$^{~~}$$^{~~}$ [.002] & .51$^{~~}$$^{~~}$$^{~~}$ [.001] & .36$^{~~}$$^{~~}$$^{~~}$ [.001] && .00 [.000]$^{~~}$ & .00 [.000]$^{~~}$ & .01 [.000]$^{~~}$\\

\hline

\multirow{3}{*}{\rotatebox[origin=c]{270}{\textbf{SQP}}}

& Random-SVM$^{r}$ & .46$^{~~}$$^{~~}$$^{~~}$ [.033] & .39$^{~~}$$^{~~}$$^{~~}$ [.002] & .27$^{~~}$$^{~~}$$^{~~}$ [.019] && .03 [.021]$^{~~}$ & .03 [.020]$^{~~}$ & .10 [.058]$^{~~}$\\

& ERisk-SVM$^{r}$ & .56$^{~~}$$^{~~}$$^{~~}$ [.018] & .49$^{~~}$$^{~~}$$^{~~}$ [.040] & .35$^{~~}$$^{~~}$$^{~~}$ [.018] && .01 [.012]$^{~~}$ & .01 [.006]$^{~~}$ & .02 [.010]$^{~~}$\\

& \textbf{ERisk-Cosine} & .70$^{\vartriangle\uparrow}$ [.012] & .58$^{\vartriangle\uparrow}$ [.009] & .40$^{\vartriangle\uparrow}$ [.003] && .01 [.010] & .01 [.006] & .02 [.006]\\
\hline

\multirow{4}{*}{\rotatebox[origin=c]{270}{\textbf{Oracles}}}
& Best conf. & .57 & .49 & .35 && .00 & .01 & .02\\
& Oracle20Random & .76 & .60 & .39 && .00 $^{~~}$ & .00 $^{~~}$ & .01$^{~~}$\\
& Oracle20ERisk & .73 & .60 & .41 && .00$^{~~}$ & .00$^{~~}$ & .00$^{~~}$\\
& Oracle & .93 & .78 & .52 && .00 & .00 & .00\\
\end{tabular}\label{tab:MSMARCO-dev_rbp_10}}
\end{table}

To compare with deep judgments, we also provide the results using the RBP metric on the TREC78 collection (see lower-part of Table~\ref{tab:MSMARCO-dev_rbp_10}) which has deep relevance judgments. We report the results using baselines, SQP methods, and Oracles. From the right side and lower-part of  Table~\ref{tab:MSMARCO-dev_rbp_10}, we can observe the residual error (RBP$_{+Error}$) estimate. As expected the residual error for TREC78 is much smaller than for MS MARCO. We can also observe that is it robust over the different trials. 
This means that for the TREC78 collection, the judgments are already deep, and the results would not be changed drastically even if we collect more judged documents for the TREC78 collection. Since the residual error estimate is low across different methods, we can say that the learning-based methods do not filter out the unjudged documents but bring  relevant documents into the retrieved list. We can also observe that $E_{Risk}$ has a lower residual than when the configurations are randomly chosen.

\section{Conclusion}
\label{sec:conclusion}
In this paper, we developed a risk-based approach for selective query processing in information retrieval. From a large number of different search configurations where both search and query expansion components and their hyperparameters vary, the model first learns from past queries which configurations are worth keeping. This selection is based on an objective function. We considered two different objective functions: $E_{Risk}$ focuses on the overall effectiveness; $N_{Risk}$ focuses on the number of queries for which the system should obtain good performance. Increasing the number of queries that are enhanced is of importance from a user point of view as it has been shown that it is easier to improve the results on easy queries to increase overall system effectiveness. It also opens a new track for system evaluation. 

The objective functions are implemented under risk functions and consider both penalty and reward. After this selection is done, the system further learns which configuration is best for which queries.  After training, the system can decide which configuration is the most appropriate to treat a new query. In addition to a machine learning based method that was used in our previous work (both classification and learning to rank was used), here we consider a Cosine  measure between trained queries and the unseen query for which the model predicts the configuration to use. 

This risk-based model has been evaluated on two tasks: ad hoc search and diversity-oriented search for which they have been proven to be effective when compared to learning to rank documents or when compared to an optimized unique configuration. The risk functions are also more effective than a random choice of configurations combined with the same selective query processing.

Selective query expansion ~\cite{amati2004query,cronen2004framework} was the first attempt toward selective query processing. 
With two configurations only, the possibilities for the system are too few to make the difference. 
In our approach we consider much more configurations and both term weighting approaches through the retrieval model component and query expansion through the QE component and its hyperparameters. This generalizes the previous attempts where either the queries were expanded ~\cite{he2007combining} or the term weighting were varying~\cite{arslan2019selective}. We show that the risk-sensitive functions we developed are both robust and effective. With our approach, on GOV2, for example, the Oracle could increase nDCG@10 by 31\% and our method achieves +26\% (Best configuration achieves 0.52, Oracle is 0.85 and our model is 0.62).  The increase using our models is $+15\%$ (resp. $+27\%$) for nDCG@10 on TREC78 (resp.  WT10G) compared to best trained. This shows that using a limited number of appropriate configurations (20 in most of the experiments) plus a selective query processing strategy on a per-query basis is effective.

In future work, we would like to conduct a qualitative data analysis to better understand which are the configurations that are selected by the $E_{Risk}$ and $N_{Risk}$ functions; how much these configurations overlap and how diverse they are. As another extension, we would like to measure the expected decrease in effectiveness when the configurations are learned on one collection and applied to another collection (transfer learning). As another extension, we would like to enhance the features used to represent training examples and test examples. Indeed, here we use a reference configuration (BM25) to calculate the (query-configuration) features. We believe that if we could rather use the genuine LETOR features associated with the query and each possible configuration, the representation would be more accurate and so will be the training. Since our risk-sensitive criteria enable us to limit the number of configurations; the extra cost would be limited.
\bibliographystyle{unsrtnat}
\bibliography{SelectiveQueryProc} 

\end{document}